\documentclass[%
reprint,
superscriptaddress,
preprintnumbers,
nofootinbib,
amsmath,amssymb,
aps,
prd,
]{revtex4-1}

\usepackage{amsmath}
\usepackage{amssymb}
\usepackage{amsthm}
\usepackage{bbm}
\usepackage{nicefrac}
\usepackage{amsfonts}

\usepackage[markup=nocolor]{changes}
\usepackage{natbib}

\usepackage{graphicx}

\usepackage[normalem]{ulem}

\usepackage{dcolumn}
\usepackage{bm}

\usepackage{color}
\usepackage[usenames,dvipsnames,svgnames,table]{xcolor}
\usepackage{slashed}
\usepackage{tensor}
\usepackage{empheq}
\newcommand{\bea}{\begin{eqnarray}}
\newcommand{\eea}{\end{eqnarray}}
\newcommand{\be}{\begin{equation}}
\newcommand{\ee}{\end{equation}}
\usepackage[normalem]{ulem}

\allowdisplaybreaks

\newcommand{\PTr}{\mathrm{P_{Tr}}}
\newcommand{\PTT}{\mathrm{P_{TT}}}

\graphicspath{{./figures/}}

\usepackage{hyperref}
\hypersetup{
	colorlinks=true,
	citecolor=blue,
	citebordercolor=red,
	linktoc=all,
	linkcolor=blue,
	urlcolor=blue
}

\def\0#1#2{\frac{#1}{#2}}

\definecolor{darkgreen}{rgb}{0,0.4,0}

\definecolor{darkgreen}{rgb}{0,0.4,0}

\usepackage{ulem}

\begin{document}
\title{Zooming in on fermions and quantum gravity}
\author{Astrid Eichhorn}
\email{eichhorn@sdu.dk}
\affiliation{CP3-Origins, University of Southern Denmark, Campusvej 55, DK-5230 Odense M, Denmark} 
\affiliation{Institut f\"ur Theoretische Physik, Universit\"at Heidelberg, Philosophenweg 16, 69120 Heidelberg, Germany}
\author{Stefan Lippoldt}
\email{s.lippoldt@thphys.uni-heidelberg.de}
\affiliation{Institut f\"ur Theoretische Physik, Universit\"at Heidelberg, Philosophenweg 16, 69120 Heidelberg, Germany}
\author{Marc Schiffer}
\email{m.schiffer@thphys.uni-heidelberg.de}
\affiliation{Institut f\"ur Theoretische Physik, Universit\"at Heidelberg, Philosophenweg 16, 69120 Heidelberg, Germany}
\begin{abstract}
We zoom in on the microscopic dynamics for fermions and quantum gravity within the asymptotic-safety paradigm.
A key finding of our study is the unavoidable presence of a nonminimal derivative coupling between the curvature
and fermion fields in the ultraviolet.
Its backreaction on the properties of the Reuter fixed point remains small for finite fermion numbers
within a bounded range. 
This constitutes a nontrivial test of the asymptotic-safety scenario for gravity and fermionic matter,
additionally supplemented by our studies of the momentum-dependent vertex flow which indicate the subleading
nature of higher-derivative couplings.
Moreover our study provides further indications that the critical surface of the Reuter fixed point
has a low dimensionality even in the presence of matter.
\end{abstract}
\maketitle
\section{Introduction}
In the search for a quantum theory of gravity that is viable in our universe, the existence of fermionic matter
must be accounted for.
Our strategy to achieve this is based on a quantum field theoretic framework that includes the metric field and fermion
fields at the microscopic level.
Such a setting requires an ultraviolet completion or extension of the effective field theory framework within which a joint
description of gravity and matter is possible up to energies close to the Planck scale.
Asymptotic safety \cite{Weinberg:1980gg,Reuter:1996cp} is the idea that scale-invariance provides a way to extend the
dynamics to arbitrarily high momentum scales without running into Landau poles which would indicate a triviality problem.
Moreover, scale-invariance is a powerful dynamical principle, that is expected to fix all but a finite number of free
parameters in an infinite dimensional space of theories.
It can be reached at a fixed point of the Renormalization Group (RG), which can be free (asymptotic freedom) or
interacting (asymptotic safety).
Compelling indications for the existence of the asymptotically safe Reuter fixed point in four-dimensional gravity have been
found, e.g., in \cite{Reuter:2001ag, Lauscher:2001ya,Litim:2003vp,Codello:2008vh, Benedetti:2009rx, Manrique:2011jc,
Becker:2014qya, Demmel:2015oqa, Gies:2016con, Denz:2016qks, Christiansen:2017bsy, Falls:2018ylp}.
For recent reviews and introductions including a discussion of open questions, see
\cite{Reuter:2012id, Ashtekar:2014kba,Eichhorn:2017egq, Percacci:2017fkn, Eichhorn:2018yfc}.

A central part of the interplay of the Standard Model with gravity is the impact of quantum gravity on the microscopic
dynamics for fermions as well as the corresponding ``backreaction'' of fermionic matter on the quantum structure
of spacetime.
In line with the observation that asymptotically safe quantum gravity could be near-perturbative
\cite{Eichhorn:2018akn,Eichhorn:2018ydy} or ``as Gaussian as it gets"
\cite{Falls:2013bv,Falls:2014tra,Falls:2017lst,Falls:2018ylp}, studies of fermion-gravity-systems
follow a truncation scheme by canonical power counting.
Furthermore, the chiral structure of the fermion sector of the Standard Model is a key guiding principle.
Thus, the leading-order terms  according to canonical power counting have been explored in the sector of chirally
symmetric fermion self interactions \cite{Eichhorn:2011pc,Meibohm:2016mkp,Eichhorn:2017eht},
and fermion-scalar interaction sector \cite{Eichhorn:2016esv,Eichhorn:2017eht}.
These are dimension-6 and dimension-8-operators, respectively. Explicitly chiral-symmetry breaking interactions,
including a mass term and two dimension-5, nonminimal couplings of fermions to gravity \cite{Eichhorn:2016vvy},
have been studied.
The effect of quantum gravity on a Yukawa coupling of fermions to scalars has been studied in
\cite{Zanusso:2009bs,Vacca:2010mj,Oda:2015sma,Eichhorn:2016esv,Hamada:2017rvn,Eichhorn:2017eht,Eichhorn:2017ylw}.
Conversely, the impact of fermionic fluctuations on the Reuter fixed point has been explored in
\cite{Dona:2012am, Dona:2013qba,Meibohm:2015twa,Eichhorn:2016vvy,Biemans:2017zca,Alkofer:2018fxj,Alkofer:2018baq}.
An asymptotically safe fixed point exists in all of these studies, as long as the fermion number is sufficiently small.
Moreover, all operators that have been explored follow the pattern that canonical dimensionality is a robust predictor
of relevance at the interacting fixed point.
Further, they confirm the conjecture that asymptotically safe quantum gravity could preserve
global symmetries \cite{Eichhorn:2017eht}, at least in the Euclidean regime.
Thus, all symmetry-breaking interactions can be set to zero consistently.
Additionally, interacting fixed points could, but need not exist for these, as in the case of the Yukawa coupling
\cite{Zanusso:2009bs,Vacca:2010mj,Oda:2015sma,Eichhorn:2016esv,Hamada:2017rvn,Eichhorn:2017eht,Eichhorn:2017ylw}.
In contrast, the interacting nature of asymptotically safe gravity percolates into the symmetric sector,
where interactions can typically not be set to zero consistently \cite{Eichhorn:2012va,Eichhorn:2017eht}.
Hence, their ``backreaction" on the asymptotically safe fixed point could be critical.
Further, this sector is a potential source of important constraints on the microscopic gravitational parameter space:
Strong gravity fluctuations could trigger new divergences in the matter sector, manifesting themselves in
complex fixed-point values for matter interactions. 
The corresponding bound on the gravitational parameter space that separates the allowed, weakly coupled gravity regime
from the forbidden strongly-coupled regime, is called the weak-gravity bound
\cite{Eichhorn:2016esv,Christiansen:2017gtg, Eichhorn:2017eht}.

In Tab.~\ref{tab:fermionresults} we provide an overview over interactions in the fermion sector that have been
explored in an asymptotically safe context.
The table contains a crucial gap, namely nonminimal, chirally symmetric interactions.
\begin{table*}[!t]
  \begin{tabular}{c|c|c|c|c|c|c}
    ref. & interaction & dimension & relevant & symmetry & free fixed point & weak-gravity bound
    \\ \hline \hline
    \cite{Eichhorn:2016vvy} & \scalebox{0.85}{$\bar{\psi}\psi$} & 3 & yes & no $\chi$ sym.& yes & no
    \\ \hline
    \cite{Eichhorn:2017eht} & \scalebox{0.85}{$\bar{\psi} \psi\, \phi$} & 4 & both possible & no $\chi$ sym. & yes & no
    \\ \hline
    \cite{Eichhorn:2016vvy} & \scalebox{0.85}{$\bar{\psi}\nabla^2\psi$} & 5 & no &  no $\chi$ sym.& yes & no
    \\ \hline
    \cite{Eichhorn:2016vvy} & \scalebox{0.85}{$R \bar{\psi}\psi$} & 5 & no & no $\chi$ sym. & yes & no
    \\ \hline
    \textbf{\emph{this work}} & \scalebox{0.85}{\boldmath{\emph{$R^{\mu\nu}\bar{\psi}\gamma_{\mu}\nabla_{\nu}\psi$}}}
     &  \textbf{\emph{6}} & \textbf{\emph{no}} & \boldmath{$\chi$} \textbf{\emph{sym.}} &  \textbf{\emph{no}}
     &  \textbf{\emph{no}}
    \\ \hline
    \cite{Eichhorn:2011pc,Meibohm:2016mkp,Eichhorn:2017eht} & \scalebox{0.85}{$(\bar{\psi}\gamma_{\mu}\gamma_5 \psi)^2$}
     & 6 & no & $\chi$ sym. & no & no
    \\ \hline
    \cite{Eichhorn:2011pc,Meibohm:2016mkp,Eichhorn:2017eht}] & \scalebox{0.85}{$(\bar{\psi}\gamma_{\mu} \psi)^2$}
     & 6 & no & $\chi$ sym. & yes & no
    \\ \hline
    \cite{Eichhorn:2016esv,Eichhorn:2017eht} &
     \scalebox{0.85}{$\bar{\psi} (\slashed{\nabla}\psi) (\partial\phi)^2$,
     $\bar{\psi} \gamma^{\mu} (\nabla_{\nu} \psi) (\partial^{\nu} \phi) (\partial_{\mu} \phi)$} & 8 & no
     & $\chi$ sym. & no & yes
  \end{tabular}
  \caption{\label{tab:fermionresults}
  We list already investigated interactions in gravity-fermion systems in order of increasing canonical dimension,
  and specify whether they are relevant at the Reuter fixed point.
  Further, we highlight that all but one interactions that respect chiral symmetry, i.e., allow a separate phase
  transformation of left-handed and right-handed fermions, are necessarily nonzero at the free fixed point.
  A subset of these exhibits a weak-gravity bound, whereas interactions that feature a free fixed point cannot
  give rise to a weak-gravity bound.
  }
\end{table*}
This is the sector that we will begin to tackle in this paper. For an analogous study in the scalar sector see
\cite{Eichhorn:2017sok}.

As our key result we find a continuation of the asymptotically safe Reuter fixed point to finite fermion numbers
that passes a nontrivial test by remaining robust under a crucial extension of the approximation to the full dynamics.
Moreover, we find further indications that the critical hypersurface of the Reuter fixed point has a low
dimensionality also in the presence of matter.

This paper is structured as follows:
In Sec.~\ref{sec:setup}, we provide an overview of the setup, and specify the approximation to the full dynamics
that we will explore in the following.
In Sec.~\ref{sec:betas} we discuss in some detail how to derive the beta functions in our setting.
In particular, we discuss the relation of the derivative expansion to the projection at finite momenta.
Sec.~\ref{sec:Nf1} provides an overview of the fixed-point results for $N_{\rm f} = 1$,
which are representative for the results at small fermion numbers.
We discuss tests of the robustness of the fixed point, the impact of the newly included nonminimal derivative interaction
on the fixed-point results in a smaller truncation, and the feature of effective universality.
Sec.~\ref{sec:wgb} contains a discussion of structural aspects of the weak-gravity bound for cubic beta functions
and highlights that no such bound exists for the nonminimal derivative interaction in the regime of gravitational
parameter space where our truncation remains viable.
In Sec.~\ref{sec:Nfgtr1} we extend our investigations to $N_{\rm f} \gg 1$, and discuss the continuation of the Reuter
fixed point to larger fermion numbers.
In Sec.~\ref{sec:conclusions} we provide a short summary of our key results and highlight possible routes forward
in gravity-matter systems in an outlook.
App.~\ref{sec:app:flow} includes a general derivation of the form of the flow equation for the dimensionless
effective action.
This form can be used to directly derive dimensionless beta functions, in contrast to the usual procedure of only
introducing dimensionless quantities after a truncation has been specified.
\section{Setup}\label{sec:setup}
The system we analyze contains a gravitational sector and a matter sector with chiral fermions.
We aim at deriving the beta functions in this system, and will employ the well-suited functional Renormalization Group.
It is based on the flow equation for the scale-dependent effective action, the Wetterich-equation
\cite{Wetterich:1992yh, Ellwanger:1993mw,Morris:1993qb},
\begin{align} \label{eq:flow}
 \dot{\Gamma}_k [\Phi ; \bar{g}] = 
 \frac{1}{2} {\rm STr} \left[
 \big( \Gamma_k^{(2)}[\Phi ; \bar{g}] + R_{k}[\bar{g}] \big)^{-1} \dot{R}_{k}[\bar{g}] \right].
\end{align}
The ``superfield'' $\Phi$ is simply a collection of all fields in our system,
 \begin{align}
 (\Phi^{A}) = \big( h_{\mu \nu}(x) , \psi^{i}(x), \bar{\psi}^{i}(x), c^{\mu}(x), \bar{c}_{\mu}(x) \big),
\end{align}
where Einsteins summation convention over the ``superindex'' $A$
contains a summation over discrete spacetime, spinor and flavor indices and
an integration over the continuous coordinates.
Here $R_k$ is a scale-dependent regulator that implements a momentum-shell wise integration of quantum fluctuations
and the dot in $\dot{\Gamma}_{k}$ refers to a derivative with respect to $t = \ln k/k_{0}$,
the RG-``time'' with $k_{0}$ an arbitrary reference scale.
The IR-regulator $R_{k}$ enters the generating functional in the form of a term that is quadratic in the fluctuation
fields and renders the Wetterich equation UV and IR finite.
Specifically, we choose a Litim-type cutoff \cite{Litim:2001up} with appropriate factors of the wave-function
renormalization for all fields.
Next to the gauge-fixing term for the metric fluctuations, it is a second source of breaking of diffeomorphism invariance.
It must be set up with respect to an auxiliary metric background $\bar{g}_{\mu\nu}$, which provides a notion of locality
and thereby enables a local form of coarse graining.
In the main part of this paper we focus on a flat background,
\begin{align}
 \bar{g}_{\mu \nu} = \delta_{\mu\nu},
\end{align}
while in this section we will keep $\bar{g}_{\mu \nu}$ arbitrary for pedagogical reasons.
For introductions and reviews of the method, see, e.g.,
\cite{Berges:2000ew, Delamotte:2007pf, Rosten:2010vm, Braun:2011pp};
specifically for gauge theories and gravity, see, e.g., \cite{Pawlowski:2005xe, Gies:2006wv,Reuter:2012id}.

The Wetterich equation provides a tower of coupled differential equations for the scale dependence of all infinitely many
couplings in theory space.
In practice, this has to be truncated to a (typically) finite-dimensional tower.
Let us briefly summarize how we proceed, before providing more details.
To construct our truncation, we define a diffeomorphism invariant ``seed action''.
Next, we expand the terms in this seed action to fifth order in metric fluctuations, defined as
\begin{align} \label{eq:lin_split}
 h_{\mu\nu} = \bar{g}_{\mu\nu}- g_{\mu\nu}.
\end{align}
This corresponds to an expansion of the seed action in vertices.
At this point all terms in the seed action, except those arising from the kinetic term for fermions,
come with one of the couplings of the seed action.
We next take into account that in the presence of a regulator and gauge fixing, the beta functions for those couplings
generically differ, when extracted from different terms.
Accordingly, we introduce a separate coupling in front of each term in the expanded action.
This provides the truncation which we analyze in the following.
To close the truncation, the couplings of higher-order vertices are partially identified with those of lower-order ones.

In more detail, these steps take the following form:
Our seed action reads
\begin{equation} \label{eq: effaction}
 S = S_{\rm grav} + S_{\rm gh} + S_{\rm mat}.
\end{equation}
Classical gravity is described by the Einstein-Hilbert action $S_{\rm EH}$,
\begin{equation}
 S_{\rm EH} = -\frac{1}{16\pi\, \bar{G}_{\rm N}} \int \!\! \mathrm{d}^{4} x \sqrt{g} \, (R - 2 \bar{\lambda}) \, .
\end{equation}
In order to tame the diffeomorphism symmetry of gravity, we choose a gauge-fixing condition $F^{\mu}$,
\begin{equation}
 F^{\mu}
 = \left( \bar{g}^{\mu \kappa} \bar{D}^{\lambda}
 - \frac{1+\beta}{4} \bar{g}^{\kappa\lambda} \bar{D}^{\mu} \right) h_{\kappa\lambda},
 \qquad \beta = 0.
\end{equation}
The gauge choice is incorporated using the gauge fixing action $S_{\rm gf}$,
\begin{equation}
 S_{\rm gf}
 = \frac{1}{32 \pi \bar{G}_{\rm N} \, \alpha} \int \!\! \mathrm{d}^{4} x \sqrt{\bar{g}} \, F^{\mu} \bar{g}_{\mu\nu} F^{\nu},
 \quad \alpha \to 0.
\end{equation}
To take care of the resulting Faddeev-Popov determinant, we use ghost fields $c^{\mu}$ and $\bar{c}_{\nu}$
with the appropriate ghost action $S_{\rm gh}$,
\begin{equation}
 S_{\rm gh}
 = \int \!\! \mathrm{d}^{4} x \sqrt{\bar{g}} \,
 \bar{c}_{\mu} \frac{\delta F^{\mu}}{\delta h_{\alpha \beta}} \mathcal{L}_{c} g_{\alpha \beta},
\end{equation}
where $\mathcal{L}_{c} g_{\alpha \beta}$ is the Lie derivative of the full metric $g_{\mu \nu}$ in ghost $c^{\mu}$ direction,
\begin{equation}
 \mathcal{L}_{c} g_{\alpha \beta}
 = 2 \bar{g}_{\rho (\alpha} \bar{D}_{\beta)} c^{\rho} + c^{\rho} \bar{D}_{\rho} h_{\alpha \beta}
 + 2 h_{\rho (\alpha} \bar{D}_{\beta)} c^{\rho}.
\end{equation}
In the following, we choose the Landau gauge, i.e., $\alpha \to 0$.
By employing a York decomposition of $h_{\mu \nu}$ we see that this choice of gauge-fixing parameters leads to
contributions from only a transverse-traceless (TT) mode $h^{\rm \scriptscriptstyle TT}_{\mu \nu}$ and a trace mode
$h^{\rm \scriptscriptstyle Tr}$,
\begin{align}
 h_{\mu \nu}
 \, \widehat{=} \, h^{\rm \scriptscriptstyle TT}_{\mu \nu} + \frac{1}{4} \bar{g}_{\mu \nu} h^{\rm \scriptscriptstyle Tr},
\end{align}
where the TT-mode satisfies $\bar{D}^{\mu} h^{\rm \scriptscriptstyle TT}_{\mu \nu} = 0$ and
$\bar{g}^{\mu\nu} h^{\rm \scriptscriptstyle TT}_{\mu \nu} = 0$,
while the trace mode is given by $h^{\rm \scriptscriptstyle Tr} = \bar{g}^{\mu \nu} h_{\mu \nu}$.
All other modes drop out of the flow equation once it is projected onto monomials with nonvanishing powers of the field.
It is important to note that the TT-mode is present in any gauge and to linear order in $h_{\mu \nu}$
a gauge invariant quantity.
Thus, for external metric fluctuations we exclusively consider the TT-mode.
For internal metric fluctuations, also the remaining trace mode is taken into account.
We summarize the purely gravitational parts of the action as $S_{\rm grav}$,
\begin{equation}\label{eq: GammaPureGrav}
 S_{\rm grav} = S_{\rm EH} + S_{\rm gf}.
\end{equation}

Next we turn to the chiral fermions.
Their minimal coupling to gravity is via the kinetic term $S_{\rm mat}^{\rm kin}$,
\begin{equation} \label{eq: Dirac}
 S_{\rm mat}^{\rm kin}
 = \sum_{i=1}^{N_{\rm f}} \int \!\! \mathrm{d}^{4} x \sqrt{g} \, \bar{\psi}^{i} \slashed{\nabla} \psi^{i} \, .
\end{equation}
For the construction of the covariant derivative for fermions, we use the spin-base invariance formalism
\cite{Gies:2013noa,Gies:2015cka,Lippoldt:2015cea}.
For our purposes, this is equivalent to using the vierbein formalism with a Lorentz symmetric gauge.
Upon expansion in $h_{\mu\nu}$, this minimal interaction between fermions and gravity gives rise to an invariant
linear in derivatives.
There are several invariants containing terms of third order in derivatives and canonical mass dimension, namely:
\begin{eqnarray} \label{eq:fermgravinvariants}
 S_{\rm mat}^{\rm \nabla^{3}} 
 & = & \sum_{i=1}^{N_{\rm f}} \int \!\! \mathrm{d}^{4} x \sqrt{g} \, \big(
 \bar{\kappa} R \bar{\psi}^i \slashed{\nabla} \psi^{i}
 + \bar{\tau} (D^{\mu} R ) \bar{\psi}^{i} \gamma_{\mu} \psi^{i}
 \\ \notag
 &{}&
 + \bar{\xi} \bar{\psi}^{i} \slashed{\nabla}^{3} \psi^{i}
 + \bar{\sigma} R^{\mu\nu}
 (\bar{\psi}^{i} \gamma_{\mu} \nabla_{\nu} \psi^{i} - (\nabla_{\nu} \bar{\psi}^{i}) \gamma_{\mu} \psi^{i}) \big),
\end{eqnarray}
where each of the invariants respects the Osterwalder-Schrader positivity of the Euclidean action.
Out of these four invariants, the ones corresponding to $\bar{\kappa}$ and $\bar{\tau}$
do not contribute linearly to an external $h^{\rm \scriptscriptstyle TT}_{\mu \nu}$,
as $R$ does not contain a transverse traceless part to linear order.
In the following, we restrict ourselves to the nonminimal coupling $\bar{\sigma}$ and neglect the $\bar{\xi}$ term.
Thus, the kinetic matter action is complemented with
\begin{equation} \label{eq:nonminmatter}
 S_{\rm mat}^{R}
 = \sum_{i=1}^{N_{\mathrm{f}}} \bar{\sigma} \! \int \!\! {\rm d}^{4} x \sqrt{g} \, R^{\mu\nu}
 \big( \bar{\psi}^{i} \gamma_{\mu} \nabla_{\nu} \psi^{i}
 - ( \nabla_{\nu} \bar{\psi}^{i} ) \gamma_{\mu} \psi^{i} \big).
\end{equation}
$S_{\rm mat}^{R}$ has all the symmetries of the original action \eqref{eq: effaction} and therefore does not enlarge
the theory space.

The nonminimal coupling $\bar{\sigma}$ introduces an invariant of cubic order in derivatives,
capturing parts of the higher-derivative structure of the fermion-gravity interaction.
Once expanded around a flat background, the interaction with $h^{\rm \scriptscriptstyle TT}_{\mu \nu}$ is given by
\begin{equation} \label{eq: invsigma}
 S_{\rm mat}^{R}
 = \sum_{i=1}^{N_{\rm f}} \bar{\sigma} \int \!\! \mathrm{d}^{4} x \,
 (\Box h^{\rm \scriptscriptstyle TT}_{\mu\nu}) \, \bar{\psi}^{i} \gamma^{\mu} \partial^{\nu} \psi^{i}
 + \mathcal{O}(h^{2}),
\end{equation}
where $\Box = - \delta^{\mu \nu} \partial_{\mu} \partial_{\nu}$ is the d'Alambertian in flat Euclidean space.
Eq.~\eqref{eq: invsigma} is the unique invariant consisting of one $h^{\rm \scriptscriptstyle TT}_{\mu \nu}$, $\psi$,
$\bar{\psi}$ and $\gamma^{\mu}$ together with two derivatives acting on the TT-mode and one derivative acting on the $\psi$.
We summarize the matter parts of the action as $S_{\rm mat}$,
\begin{equation}\label{eq: GammaMat}
 S_{\rm mat} = S_{\rm mat}^{\rm kin} + S_{\rm mat}^{R}.
\end{equation}

After having specified our complete seed action, we expand the scale-dependent effective action in powers of the fluctuation field, 
\begin{equation} \label{eq: VertExpand}
 \Gamma_{k}[\Phi; \bar{g}]
 = \sum_{n=0}^{\infty} \frac{1}{n!} \Gamma_{k \, A_{1} \ldots A_{n}}^{(n)}[0; \bar{g}] \Phi^{A_{n}} \ldots \Phi^{A_{1}},
\end{equation}
where $\Gamma_{k}^{(n)}$ refers to functional derivatives with respect to the field $\Phi$,
\begin{align}\label{eq: GN}
 \Gamma_{k \, A_{1} \ldots A_{n}}^{(n)}[\Phi;\bar{g}]
 = \Gamma_{k} [\Phi;\bar{g}]
 \frac{\overleftarrow{\delta}}{\delta \Phi^{A_{1}}} \ldots \frac{\overleftarrow{\delta}}{\delta \Phi^{A_{n}}}.
\end{align}
Note the order of the indices and fields, which is important to keep in mind for the Grassmann-valued quantities.

By using this vertex form, the flow of 5 individual couplings $\bar{\lambda}_{2}$,
$\bar{\lambda}_{3}$, $\bar{G}_{h}$, $\bar{G}_{\psi}$ and $\bar{\sigma}$ as well as the anomalous dimension of
two wave-function renormalizations $Z_{h}$ and $Z_{\psi}$ is disentangled, cf.~Tab.~\ref{tab: Couplings}
and see Sect.~\ref{sec:betas} for more details.
\begin{table}[!t]
 \begin{tabular}{c||c|c|c}
  Couplings & $S_{\rm EH}$ & $S_{\rm mat}^{\rm kin}$ & $S_{\rm mat}^{R}$
  \\ \hline \hline \vphantom{${{\big(^{A}}^{A}_{A}}_{A}$}
  $\Gamma_{k}^{(2)}$ & $\bar{\lambda}_{2}$, $Z_{h}$ & $Z_{\psi}$ & --
  \\ \hline \vphantom{${{\big(^{A}}^{A}_{A}}_{A}$}
  $\Gamma_{k}^{(3)}$ & $\bar{\lambda}_{3}$, $\bar{G}_{h}$ & $\bar{G}_{\psi}$ & $\bar{\sigma}$
  \\ \hline \vphantom{${{\big(^{A}}^{A}_{A}}_{A}$}
  $\Gamma_{k}^{(4)}$ & $\bar{\lambda}_{4} = \bar{\lambda}_{3}$, $\bar{G}_{h,4} = \bar{G}_{h}$
   & $\bar{G}_{\psi, 4} = \bar{G}_{\psi}$ & $\bar{\sigma}_{4} = \bar{\sigma}$
  \\ \hline \vphantom{${{\big(^{A}}^{A}_{A}}_{A}$}
  $\Gamma_{k}^{(5)}$ & $\bar{\lambda}_{5} = \bar{\lambda}_{3}$, $\bar{G}_{h,5} = \bar{G}_{h}$
   & $\bar{G}_{\psi, 5} = \bar{G}_{\psi}$ & $\bar{\sigma}_{5} = \bar{\sigma}$
 \end{tabular}
 \caption{
 We list the couplings and wave function renormalizations appearing in the $n$-th functional derivative $\Gamma_{k}^{(n)}$
 of the effective action and indicate to which part of the seed action, $S_{\rm EH}$, $S_{\rm mat}^{\rm kin}$
 or $S_{\rm mat}^{R}$ they are related.
 }
 \label{tab: Couplings}
\end{table}
Here the barred couplings, e.g., $\bar{G}_{\psi}$ and $\bar{G}_h$, refer to dimensionful couplings.

\begin{figure}[!t]
  \includegraphics[width=\linewidth]{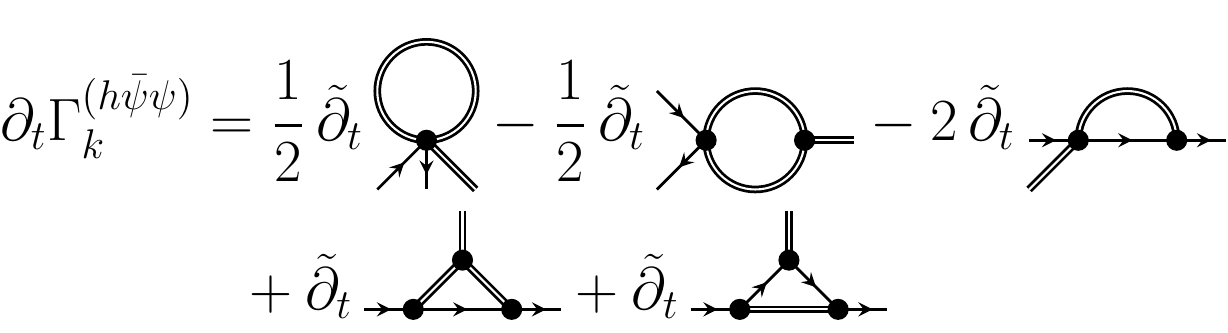}
  \caption{\label{fig:flowdiags}
  We show the diagrams contributing to the flow of the gravity-fermion system.
  Double lines denote metric fluctuations, single lines fermions.
  Each diagram is understood to carry a regulator insertion on one of the propagators, that $\tilde{\partial_t}$ acts upon.
  When $\tilde{\partial_{t}}$ is evaluated, a diagram with  $n$ internal propagators becomes the sum of $n$
  diagrams, such that the regulator insertion occurs on each of the internal propagators once.
  }
\end{figure}

For the gravity-fermion vertex the contributing diagrams are shown in Fig.~\ref{fig:flowdiags}.
This highlights the necessity to truncate the tower of vertices, as the flow of each $n$-point vertex depends on the
$(n + 1)$- and $(n + 2)$-point vertices.
We use the seed action in Eq.~\eqref{eq: effaction} to parametrize the vertices appearing in the diagrams.
When generating, e.g., a graviton three-point vertex or a graviton four-point vertex for the scale-dependent effective action from the seed
action  by expanding to the appropriate power in $h_{\mu\nu}$, both would depend on the same Newton coupling $\bar{G}_{\rm N}$ and the same cosmological
constant $\bar{\lambda}$ due to diffeomorphism symmetry.
However, the gauge fixing and the regulator break diffeomorphism symmetry.
Hence, the effective action is known to satisfy Slavnov-Taylor identities instead,
\cite{Ellwanger:1995qf,Reuter:1996cp,Pawlowski:2005xe,Pawlowski:2003sk,Manrique:2009uh,Donkin:2012ud}.
As these identities in general are much more involved, there is no such simple relation between the three- and four-point
vertex of the effective action as there is for the seed action.
In other words, the breaking of diffeomorphism symmetry leads to an enlargement of theory space in which the couplings parameterizing the vertices are independent. There are different routes towards a truncation of this large theory space. In principle, one could pick some random tensor structure and momentum-dependence in each $n-$point function and parameterize this by some coupling. Then, the connection to the diffeomorphism-invariant seed action would be lost completely. Instead, we derive the tensor structures of the vertices from the seed-action, but also take into account that the various couplings are now independent.
Specifically, we proceed using the following recipe:
The structure of the $n$-point vertex is drawn from the seed action,
\begin{align}
 \label{eq: VertexGrav}
 \Gamma_{k \, A_{1} \ldots A_{n}}^{(n)}
 = Z^{\frac{1}{2} B_{1}}_{\Phi \,\,\,\,\, A_{1}} \ldots Z^{\frac{1}{2} B_{n}}_{\Phi \,\,\,\,\, A_{n}}
 S^{(n)}_{B_{1} \ldots B_{n}} \Big|_{\bar{\lambda} \to \bar{\lambda}_{n}} ,
\end{align}
where the replacement $\bar{\lambda} \to \bar{\lambda}_{n}$ only affects pure gravity vertices.
Furthermore in Eq.~\eqref{eq: VertexGrav} the metric fluctuations of the purely gravitational action $S_{\rm grav}$
are rescaled according to
\begin{equation} \label{eq: PureGravResc}
 S_{\rm grav}: \quad
 (h^n)_{\mu\nu} \to (16 \pi)^{\frac{n}{2}} \bar{G}_{\rm N} ( \bar{G}_{h} )^{\frac{n}{2}-1} (h^n)_{\mu\nu} ,
\end{equation}
whereas the graviton in $S_{\rm gh}$ and $S_{\rm mat}$ is rescaled to
\begin{align} \label{eq: GhostResc}
S_{\rm gh}:{}& \quad (h^n)_{\mu\nu} \to (16 \pi)^{\frac{n}{2}} ( \bar{G}_{h} )^{\frac{n}{2}} (h^n)_{\mu\nu} ,
 \\ \label{eq: MinMattResc}
 S_{\rm mat}:{}& \quad (h^n)_{\mu\nu} \to (16 \pi)^{\frac{n}{2}} ( \bar{G}_{\psi} )^{\frac{n}{2}} (h^n)_{\mu\nu} .
\end{align}
This rescaling breaks diffeomorphism symmetry and helps us choosing a basis in the appropriate theory space. Note that the field-redefinitions in Eq.~\eqref{eq: PureGravResc}, \eqref{eq: GhostResc} and \eqref{eq: MinMattResc} are not to be understood as actual field-redefinitions in the effective action. They are just a way of arriving at a parameterization of the truncated effective action in the enlarged theory space.\\
In the following we use the term ``avatar'', when a single coupling in the seed action leads to various incarnations
in the effective action, e.g., $\bar{G}_{h}$ and $\bar{G}_{\psi}$ are avatars of the Newton coupling $\bar{G}_{\rm N}$.

In order to close the flow equation, we identify couplings of higher order $n$-point vertices with the corresponding
couplings of the three-point vertex.
This was already implicitly done with the rescaling
in Eqs.~\eqref{eq: PureGravResc}, \eqref{eq: GhostResc} and \eqref{eq: MinMattResc}
and with the usage of one single coupling $\bar{\sigma}$.
Similarly, all $n \geq 3$-point vertices arising from the cosmological-constant part of the seed action are parametrized by
one coupling $\bar{\lambda}_3$, i.e., $\bar{\lambda}_{4} \to \bar{\lambda}_{3}$
and $\bar{\lambda}_{5} \to \bar{\lambda}_{3}$.
The relation between $\bar{\lambda}_{2}$ and the gravitational mass-parameter $\bar{\mu}_{h}$
that is often used in the literature, reads $-2 \bar{\lambda}_{2} = \bar{\mu}_{h}$.
In the next section we provide details on how the beta functions are extracted from the sum of the diagrams in Fig.~\ref{fig:flowdiags}.
\section{How to obtain beta functions} \label{sec:betas}
We now discuss in some detail how to derive beta functions.
We concentrate on the dimensionless couplings, which are obtained from their dimensionful counterparts by
a multiplication with an appropriate power of $k$.
Dimensionful couplings are denoted with overbars, e.g., $\bar{G}_{\psi}, \bar{\lambda}_{2}$ etc.,
whereas their dimensionless counterparts lack the overbar, e.g., $G_{\psi}, \lambda_{2}$ etc.

A key goal of ours is to test the quality of our truncation.
Thus, we place a main focus on the momentum-dependence of the flow,
i.e., the dependence of the $n$-point vertices on the momenta of the fields.
Higher-order momentum-dependencies than those included in the truncation are in general present.
This implies that different projection schemes might yield different results when working in truncations.
We will discuss these different schemes and their relation to each other in the following.
\subsection{Fermionic Example}
As a concrete example let us consider the fermionic sector.
To arrive at beta functions, we have to take several steps.
First we define a projector $\mathbbm{P}^{(3)}_{p_{2},p_{3}}$ on the gravity-fermion vertex.
Its form is motivated by the tensor structure of the considered three-point function,
\begin{align} \label{eq:Gamma_Psi_Trunc}
 S_{\rm mat}^{\rm kin} {}& \big[
 g \! = \! \delta \! + \! ( 16 \pi \bar{G}_{\psi} )^{\frac{1}{2}} h^{\rm \scriptscriptstyle TT} ,
 \psi, \bar{\psi} \big]
 \\ \notag
 {}& \!\! =
 \sum\limits_{i=1}^{N_{\rm f}} \int \!\! \mathrm{d}^{4} x \big[ \bar{\psi}^{i} \slashed{\partial} \psi^{i}
 - 2 \pi^{\frac{1}{2}} \bar{G}_{\psi}^{\frac{1}{2}}
 h^{\! {\scriptscriptstyle \rm TT}}_{\! \mu \nu} \bar{\psi}^{i} \gamma^{\mu} \partial^{\nu} \! \psi^{i} \big]
 + \mathcal{O}(h^{2}).
\end{align}
By taking the corresponding functional derivatives of Eq.~\eqref{eq:Gamma_Psi_Trunc} and evaluating in momentum space,
while using the projector onto transverse traceless symmetric tensors $\Pi^{\scriptscriptstyle \rm TT}$, we find that,
\begin{align}
 {}& \int\limits_{x,y,z} \!\!\!
 e^{i (p_{1} \cdot x + p_{2} \cdot y + p_{3} \cdot z)} \,
 S_{\rm mat}^{\rm kin}
 \frac{\overleftarrow{\delta}}{\delta h^{\scriptscriptstyle \rm TT}_{\mu \nu}(x) \! }
 \frac{\overleftarrow{\delta}}{\delta \psi^{i}(y) \!}
 \frac{\overleftarrow{\delta}}{\delta \bar{\psi}^{j}(z)}
 \\ \notag
 {}& = (2 \pi)^{4} \delta(p_{1} \! + \! p_{2} \! + \! p_{3}) (- i 2 \pi^{\frac{1}{2}} \bar{G}_{\psi}^{\frac{1}{2}})
 \Pi^{{\scriptscriptstyle \rm TT} \, \, \mu \nu}_{\, \rho \sigma}(p_{1}) \gamma^{\rho} p_{2}^{\sigma} \, \delta_{i j}.
\end{align}
Of the three momenta, only two are independent, the third can be eliminated by momentum conservation.
Thus we define the projector $\mathbbm{P}_{p_{2},p_{3}}^{(3)}$ on $\bar{G}_{\psi}$ as
\begin{align} \notag
 \mathbbm{P}_{p_{2},p_{3} \, \mu \nu}^{(3) \quad i j} {}&(x,y,z)
 \\ \label{eq: Proj3}
 {}& \hspace{-0.9cm}
 = \frac{i \gamma_{\rho} \, p_{2\sigma}}{10 \pi^{\frac{1}{2}} N_{\rm f} p^{2}}
 \Pi^{{\scriptscriptstyle \rm TT} \,\, \rho \sigma}_{\, \mu\nu} \! (p_{1})
 \, e^{i ( p_{2} \cdot y + p_{3} \cdot z )}
 \delta(x) \delta^{i j},
\end{align}
which we evaluate at the symmetric point for the momenta, $p_{1}^{2} = p_{2}^{2} = -2 (p_{1} \cdot p_{2}) = p^{2}$.
The normalization of $\mathbbm{P}^{(3)}_{p_{2},p_{3}}$ follows from
\begin{align}
 \Pi^{{\scriptscriptstyle \rm TT} \, \, \mu \nu}_{\, \mu \sigma}(p_{1})
 p_{2}^{\sigma} p_{2 \, \nu}
 = \frac{5}{3} \left( p_{2}^{2} - \tfrac{(p_{1} \cdot p_{2})^{2}}{p_{1}^{2}} \right).
\end{align}
Using $\mathbbm{P}^{(3)}_{p_{2}, p_{3}}$ we define the projected dimensionful vertex $\bar{V}$ as
\begin{align} \label{eq:def_Vbar}
 \bar{V}(p^{2})
 =& \!\!\! \int\limits_{x,y,z} \!\!\!
 {\rm tr} \bigg[
 \mathbbm{P}^{(3) \quad i j}_{p_{2},p_{3} \, \mu \nu}(x,y,z)
 \Gamma_{k}
 \tfrac{\overleftarrow{\delta}}{\delta h_{\mu \nu}(x) \! }
 \tfrac{\overleftarrow{\delta}}{\delta \psi^{i}(y) \!}
 \tfrac{\overleftarrow{\delta}}{\delta \bar{\psi}^{j}(z)}
 \bigg]_{\Phi = 0},
\end{align}
where $\rm tr$ implies the trace over Dirac and flavor indices.
This definition is independent of any truncation, while a truncation for $\bar{V}$ can be viewed as choosing
a specific point in theory space.
For instance, when evaluating $\bar{V}$ for our chosen truncation we find that $\bar{V}$ is equal to
$\sqrt{\bar{G}_{\psi}} (1 - 2 \bar{\sigma} p^{2})$.
Having defined $\bar{V}$, we aim at deriving the beta function for the dimensionless counterpart $V$,
\begin{align} \label{eq:def_V_dimless}
 V( \tfrac{p^{2}}{k^{2}} ) = \frac{k}{Z_{h}^{\frac{1}{2}}(p^{2}) Z_{\psi}(p^{2})} \bar{V}( p^{2} ).
\end{align}
Note that $\bar{V}$ carries a non-trivial dimension,
as the gravity-fermion vertex contains an additional momentum $p$.
The scale derivative of $V$ reads
\begin{align} \notag
 \beta_{V}(\tfrac{p^{2}}{k^{2}})
 ={}& V( \tfrac{p^{2}}{k^{2}} ) \big( 1 + \eta_{\psi}(p^{2}) + \tfrac{1}{2} \eta_{h}(p^{2}) \big)
 + 2 \tfrac{p^{2}}{k^{2}} V'(\tfrac{p^{2}}{k^{2}})
 \\
 {}& + k \dot{\bar{V}}(p^{2}),\label{eq:betaV}
\end{align}
where one has to take into account the scaling of the momentum, $p \sim k$.
Here $\eta_{h}$ and $\eta_{\psi}$ are the anomalous dimensions,
\begin{align}
 \eta_{h}(p^{2}) = - \frac{\dot{Z}_{h}(p^{2})}{Z_{h}(p^{2})},
 \qquad 
 \eta_{\psi}(p^{2}) = - \frac{\dot{Z}_{\psi}(p^{2})}{Z_{\psi}(p^{2})}.
\end{align}
We can read off $\dot{\bar{V}}$ by replacing $\Gamma_{k}$ with $\dot{\Gamma}_{k}$ in equation \eqref{eq:def_Vbar},
\begin{align} \label{eq:dotV}
 \dot{\bar{V}}(p^{2}) = \tfrac{1}{k} {\rm Flow}^{(3)}_{\psi}(p^{2}),
\end{align}
where ${\rm Flow}^{(3)}_{\psi}$ is a short hand for the contributing diagrams in Fig.~\ref{fig:flowdiags},
\begin{align} \notag
 {\rm Flow}^{(3)}_{\psi}(p^{2})
 \! ={}& \! \tfrac{k}{2} \!\!\!\!\! \int\limits_{x,y,z} \!\!\!\!\! {\rm tr} \! \bigg[ \! {\rm STr} \Big[
 \tfrac{1}{\Gamma_{k}^{(2)} \! + \! R_{k}} \dot{R}_{k}  \Big] \!
 \tfrac{\overleftarrow{\delta}}{\delta h_{\mu \nu}(x) \! }
 \tfrac{\overleftarrow{\delta}}{\delta \psi^{i}(y) \!}
 \tfrac{\overleftarrow{\delta}}{\delta \bar{\psi}^{j}(z)} \big|_{\Phi = 0}
 \\ \label{eq:Flowdef}
 {}& \times \mathbbm{P}^{(3) \quad i j}_{p_{2},p_{3} \, \mu \nu}(x,y,z) \bigg].
\end{align}
Here we made use of Eq.~\eqref{eq:flow}.

By inserting the expression for $\dot{\bar{V}}$ given in Eq.~\eqref{eq:dotV} into Eq.~\eqref{eq:betaV} for $\beta_V$
we finally arrive at the beta function $\beta_{V}$ for $V$,
\begin{align} \label{eq: vertflow}
 \beta_{V}(\tfrac{p^{2}}{k^{2}})
 ={}& {\rm Flow}^{(3)}_{\psi}(p^{2})
 \\ \notag
 {}& + V(\tfrac{p^{2}}{k^{2}}) \big( 1 + \eta_{\psi}(p^{2}) + \tfrac{1}{2} \eta_{h}(p^{2}) \big)
 + 2 \tfrac{p^{2}}{k^{2}} V'(\tfrac{p^{2}}{k^{2}}) .
\end{align}
This equation will take center stage in our analysis of the momentum-dependence of the flow and tests of robustness
of the truncation.
We highlight that in general the right-hand-side of the flow equation generates terms beyond the chosen truncation.
In Eq.~\eqref{eq: vertflow} the consequence is, that our truncation does not capture the full momentum-dependence that is
generated.
Accordingly, the fixed-point equation $\beta_V(\tfrac{p^2}{k^2}) = 0$ cannot be satisfied
for all momenta, but instead only at selected points.
We will extensively test how large the deviations of $\beta_V$ from zero are in order to judge the quality of different
truncations.
\subsection{Projection schemes}
We perform our analysis in several different projection schemes, as a comparison between the fixed-point structure of
the different truncations provides indications for or against the robustness of the fixed point.
We now motivate the use and explain the details of these three projection schemes.

Using Eq.~\eqref{eq: vertflow}, the momentum-dependent fixed-point vertex $V^{\ast}(\tfrac{p^{2}}{k^{2}})$
could be found by demanding $\beta_{V}(\tfrac{p^{2}}{k^{2}}) = 0$ and solving Eq.~\eqref{eq: vertflow}.
In practice, we choose an ansatz $V_{\rm trunc}(\tfrac{p^{2}}{k^{2}})$ for $V(\tfrac{p^{2}}{k^{2}})$,
which is part of choosing a truncation.
At a point in theory space defined by $V_{\rm trunc}$, Eq.~\eqref{eq: vertflow} holds, but indicates that
terms not yet captured by $V_{\rm trunc}$ are generated.
These are present in Eq.~\eqref{eq: vertflow}, so that we need to truncate
the beta function $\beta_{V} \to \beta_{V}^{\rm trunc}$ in order to close the system.
For example, in our setup, we restrict $V_{\rm trunc}$ to a polynomial up to first order in $\tfrac{p^{2}}{k^{2}}$, i.e.,
\begin{equation} \label{eq: Ansatz}
 V_{\rm trunc}(\tfrac{p^{2}}{k^{2}})
 = \sqrt{G_{\psi}} - 2 \sigma_{\rm mod} \tfrac{p^{2}}{k^{2}},
\end{equation}
and
\begin{equation} \label{eq: betaAnsatz}
 \beta_{V}^{\rm trunc}(\tfrac{p^{2}}{k^{2}})
 = \frac{1}{2 \sqrt{G_{\psi}}} \beta_{G_{\psi}} - 2 \beta_{\sigma_{\rm mod}} \tfrac{p^{2}}{k^{2}}.
\end{equation}
Here we introduced a modified version of the coupling $\sigma$,
\begin{align}\label{eq:redef}
 \sigma_{\rm mod} = \sqrt{G_{\psi}} \sigma,
\end{align}
where $G_{\psi}$ and $\sigma$ are the dimensionless counterparts of $\bar{G}_{\psi}$ and $\bar{\sigma}$,
\begin{align}
 G_{\psi} = k^{2} \bar{G}_{\psi}, \quad
 \sigma = k^{2} \bar{\sigma}.
\end{align}
However, this specific ansatz does not satisfy Eq.~\eqref{eq: vertflow} for all values of $\tfrac{p^{2}}{k^{2}}$.
Accordingly, the right-hand side of Eq.~\eqref{eq: vertflow} differs from Eq.~\eqref{eq: betaAnsatz}.
This is simply an example for the general fact that, plugging a truncation into the right-hand-side
of the Wetterich equation, terms beyond the truncation are generated and therefore the truncation is not closed.

As $\beta_{V}^{\rm trunc}$ is not equal to $\beta_{V}$ for all momenta, we can choose selected points
in the interval $\tfrac{p^{2}}{k^{2}} \in [0,1]$ for which we demand that $\beta_{V}^{\rm trunc}$ is exactly equal to
$\beta_{V}$ at these points, see, e.g., Eqs.~\eqref{eq: betan=2} and \eqref{eq: BSigMod}.
However, we can also choose superpositions of more values for $\tfrac{p^{2}}{k^{2}}$, see, e.g.,
Eq.~\eqref{eq: vertflowg} for $\sigma=0$.
Even though this superposition might lead to $\beta_{V}^{\rm trunc}$ being not exactly equal to $\beta_{V}$ at any point,
it can still lead to an overall better description of the full momentum dependence, by being almost equal in a larger region.
The values of the coefficients in the ansatz, i.e., $\sqrt{G_{\psi}}$ and $\sigma_{\rm mod}$, depend on this choice.

Let us now compare two popular choices, namely the derivative expansion about $\tfrac{p^{2}}{k^{2}} = 0$,
and a projection at various values for $\tfrac{p^{2}}{k^{2}}$.
Working within a derivative expansion about $\tfrac{p^{2}}{k^{2}} = 0$,  one extracts the flow of the $n$-th coefficient
of the polynomial by the $n$-th derivative of Eq.~\eqref{eq: vertflow}, evaluated at $\tfrac{p^{2}}{k^{2}} = 0$.
Specifically, for the chosen ansatz in Eq.~\eqref{eq: Ansatz} together with Eq.~\eqref{eq: betaAnsatz}, this yields:
\begin{align}
 \beta_{G_{\psi}}^{\rm DE} = 2 \sqrt{G_{\psi}} \beta_{V}(0),
 \qquad
 \beta_{\sigma_{\rm mod}}^{\rm DE}
 = - \frac{1}{2} \beta_{V}'(0).
\end{align}
This expansion ensures that $\beta_{V}$ and its derivative are equal to $\beta_{V}^{\rm trunc}$
and its derivative at $\tfrac{p^{2}}{k^{2}} = 0$.
However, the derivative expansion to this order does not satisfy this equality away from $\tfrac{p^{2}}{k^{2}} = 0$.
This simply means that higher-order terms in the derivative expansion around $\tfrac{p^{2}}{k^{2}} = 0$
are generated by the flow.
By the evaluation at a single point in $\tfrac{p^{2}}{k^{2}}$, this scheme is very sensitive to local fluctuations
at $\tfrac{p^{2}}{k^{2}} =0$, which might cause deviations for larger momenta.

Alternatively,  we can choose finite momenta, e.g., $\tfrac{p^{2}}{k^{2}} = 1$, to extract one of the beta functions.
Equating $\beta_{V}$ and $\beta_{V}^{\rm trunc}$ at $\tfrac{p^{2}}{k^{2}} = \tfrac{1}{2}$ and $\tfrac{p^{2}}{k^{2}} = 1$,
and solving for the beta functions yields
\begin{align}\label{eq: betan=2}
 \beta_{G_{\psi}}^{(0,1)}
 ={}& 2 \sqrt{G_{\psi}} \big( 2 \beta_{V}(\tfrac{1}{2}) - \beta_{V}(1) \big),
 \\ 
 \beta_{\sigma_{\rm mod}}^{(0,1)}
 ={}& \beta_{V}(\tfrac{1}{2}) - \beta_{V}(1).
 \label{eq: BSigMod}
\end{align}
In this scheme, the beta functions $\beta_{V}$ and $\beta_{V}^{\rm trunc}$ by construction
are equal at $\tfrac{p^{2}}{k^{2}} = \tfrac{1}{2}$ and $\tfrac{p^{2}}{k^{2}} = 1$.
Thus, it provides an interpolation between these momenta, while the derivative expansion provides an extrapolation
from $\tfrac{p^{2}}{k^{2}} = 0$ onwards.
The same projection schemes can analogously be applied to other $n$-point functions, including the anomalous dimensions.
We will refer to the projection at $n$ different values for the momentum $\tfrac{p^{2}}{k^{2}}$
by \emph{$n$-sample-point projection} in the following.
More specifically, starting from Eq.~\eqref{eq: betan=2}, the beta functions for $G_{\psi}$ and $\sigma$ take the
following form
\begin{align}
 \notag
 \beta_{G_{\psi}}
 ={}& 2 \sqrt{G_{\psi}} \big[ 2 C(\tfrac{k^{2}}{2}) V_{\rm trunc}(\tfrac{1}{2}) - C(k^{2}) V_{\rm trunc}(1)
 \\ \label{eq: vertflowg}
 {}& \hspace{1.2cm} + 2 {\rm Flow}^{(3)}_{\psi}(\tfrac{k^{2}}{2}) - {\rm Flow}^{(3)}_{\psi}(k^{2}) \big] ,
 \\ \nonumber
 \beta_{\sigma}
 ={}& 2 \, \sigma + \frac{1}{2 \sqrt{G_{\psi}}} \Bigg( V_{\rm trunc}(1) \left( - C_{1}(k^{2})
 + \frac{\beta_{G_{\psi}}}{2 G_{\psi}} \right)
 \\ \label{eq: vertflowsigma}
 {}& \quad\quad \quad\quad\quad\quad\quad-{\rm Flow}^{(3)}_{\psi}(k^{2}) \Bigg) ,
\end{align}
with
\begin{eqnarray}
 C(p^{2})
 & = & 1 + \tfrac{1}{2} \eta_{h}(p^{2}) + \eta_{\psi}(p^{2}).
\end{eqnarray}
In practice, the ingredients to evaluate the beta functions are the following:
${\rm Flow}_{\psi}^{(3)}$ is given by the sum of diagrams in Fig.~\ref{fig:flowdiags}
which uses xAct \cite{Brizuela:2008ra,Brizuela:2008ra,2008CoPhC.179..597M,2007CoPhC.177..640M,2008CoPhC.179..586M}
as well as the FORM-tracer \cite{Cyrol:2016zqb},
$V_{\rm trunc}$ is given by Eq.~\eqref{eq: betaAnsatz} and the anomalous dimensions are extracted from a projection
of the corresponding two-point functions at
$\tfrac{p^{2}}{k^{2}} = 1$, as in \cite{Christiansen:2014raa,Christiansen:2015rva,Meibohm:2015twa,Denz:2016qks}.

We now provide our motivations for using projections with $n = 1$ and $n = 2$ sampling points.
The derivative expansion at $\tfrac{p^{2}}{k^{2}} = 0$ for the gravity-matter avatars of the Newton coupling does
not capture all properties of the flow in a quantitatively reliable way cf.~the discussion in \cite{Eichhorn:2018akn}.
In particular, a derivative expansion of the Einstein-Hilbert truncation at $p^{2} = 0$, together with a
momentum-independent anomalous dimension for the graviton results in a slightly screening property of gravity
fluctuations on the Newton coupling.
We expect that at higher orders in the truncation, the derivative expansion becomes quantitatively reliable,
but in our truncation projections at finite momenta are preferable.

\begin{figure}[!t]
 \centering
 \includegraphics[width=0.5\linewidth]{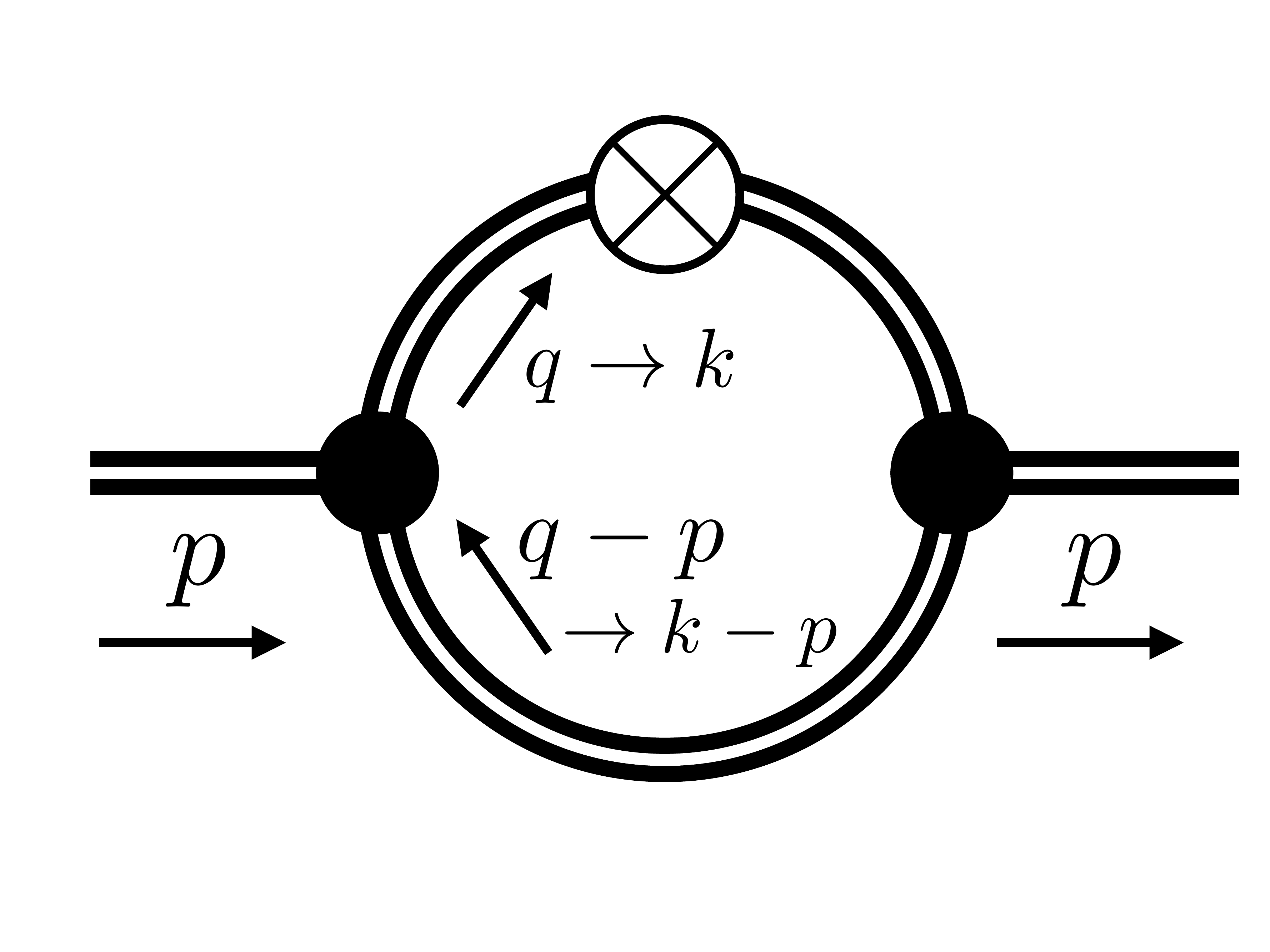}
 \caption{\label{fig:mompeak}
 The regulator insertion in the flow equation results in $q^2 \approx k^2$ from the loop integral,
 indicated by the arrows in the diagram.
 }
\end{figure}

Instead, an expansion at finite momenta is expected to be more stable in small truncations.
This is easiest to appreciate when thinking of the flow equation in terms of a vertex expansion:
The $n$-point functions that enter the flow depend on $n-1$ momenta.
Of these, one becomes the loop momentum in the flow equation.
Due to the properties of the regulator, the momentum integral over the loop momentum is peaked at $q^{2} \approx k^{2}$.
Accordingly, the flow depends on the vertex at a finite momentum, not vanishing momentum, cf.~Fig.~\ref{fig:mompeak}.

Accordingly, a good approximation of the full flow might require higher orders in the derivative expansion
around $\tfrac{p^{2}}{k^{2}} = 0$ than in projection schemes at finite momentum.
For technical simplicity, a symmetric point where the magnitudes of all momenta at the vertex are chosen to be
the same nonzero value is preferable, although the example in Fig.~\ref{fig:mompeak} showcases that a non-symmetric
point is likely to most accurately capture the momentum-dependence of the vertex as it is relevant for the feedback into
the flow equation.

We point out that for this type of projections, a one-to-one mapping between the couplings extracted in this way
and the couplings of the action written in a derivative expansion in terms of curvature invariants, as it is usually done,
becomes more involved.
For the derivative expansion about $\tfrac{p^2}{k^2}=0$, this mapping is one-to-one.
Specifically, projecting onto a $\tfrac{p^{2}}{k^{2}}$ term at finite $\tfrac{p}{k}$ yields a different result
than projection at vanishing $\tfrac{p}{k}$.
This remains the case even in untruncated theory space, where the couplings in a derivative expansion around
zero momentum and the couplings in a projection at finite momenta satisfy a nontrivial mapping onto each other.
In an untruncated theory space, such a difference in the choice of basis does not matter for the universal properties of
the fixed point.
In truncations, such choices can make a difference, as some expansions are better suited to capturing the flow already
in small truncations.
One might tentatively interpret the results in the Einstein-Hilbert truncation and small extensions thereof
\cite{Christiansen:2014raa, Christiansen:2015rva, Meibohm:2015twa,Denz:2016qks,Eichhorn:2018akn,Christiansen:2017cxa,
Eichhorn:2018ydy} as implying that projections with $n$ sampling points are preferred over the derivative expansion about
vanishing momentum.

For the fermion-gravity vertex, we consider the following three approximations
\begin{itemize}
 \item $n\!\!=\!\!1$ projection:
 We set $\sigma=0$ in $V_{\rm trunc}$ (Eq.~\eqref{eq: Ansatz}) and $\beta_{\sigma_{\rm mod}}=0$ in $\beta_{V}^{\rm trunc}$
 (Eq.~\eqref{eq: betaAnsatz}) and project onto $\beta_{G_{\psi}}$ using the projection point $\tfrac{p^2}{k^2}=1$.
 In analogous systems, such a projection has been called bilocal,
 \cite{Christiansen:2017cxa, Eichhorn:2018akn, Eichhorn:2018ydy}.
 \item $n\!\!=\!\!2$ with $\sigma=0$ projection:
 We set $\sigma=0$ in $V_{\rm trunc}$ (Eq.~\eqref{eq: Ansatz}) and $\beta_{\sigma_{\rm mod}}=0$ in Eq.~\eqref{eq: betaAnsatz}
 and project onto $\beta_{G_{\psi}}$ using the two projection points $\tfrac{p^2}{k^2}=1$ and
 $\tfrac{p^2}{k^2}=\tfrac{1}{2}$, as in Eq.~\eqref{eq: vertflowg}.
 \item $n\!\!=\!\!2$ with $\sigma \neq 0$ projection:
 This projection contains an extension of the truncation by $\beta_{\sigma}$, and uses Eq.~\eqref{eq: vertflowg}
 and Eq.~\eqref{eq: vertflowsigma}, based on the full expressions for  $V_{\rm trunc}$ (Eq.~\eqref{eq: Ansatz}) and
 $\beta_{V}^{\rm trunc}$ (Eq.~\eqref{eq: betaAnsatz}).
\end{itemize}

For completeness let us explain how we extract the remaining gravitational couplings,
$\bar{\lambda}_{2}$, $\bar{\lambda}_{3}$, $\bar{G}_{h}$, and the wave-function renormalizations, $Z_{h}$, $Z_{\psi}$.
An analogous parameterization to Eqs.~\eqref{eq:def_Vbar} and \eqref{eq: Ansatz} holds for the two-fermion, as well as
the two- and three-graviton vertices.
Thus, in all approximation schemes under consideration, we use a projection with $n\!\!=\!\!2$ sample-points
of $\Gamma_{k}^{(2)}$ and $\Gamma_{k}^{(3)}$ at $p^{2} = 0$ and $p^{2} = k^{2}$ for these couplings,
i.e., we define them as follows:
\begin{widetext}
\begin{align} \notag
 Z_{h}(p^{2}) ={}& \mathcal{N}_{Z_{h}}(p^{2}) \! \int\limits_{x,y} \! \Big[
 \Gamma_{k}[\Phi;\bar{g} = \delta] \tfrac{\overleftarrow{\delta}}{\delta h^{\mu \nu}(x)}
 \tfrac{\overleftarrow{\delta}}{\delta h^{\rho \sigma}(y)} \Big]_{\Phi = 0}
 \,\, \delta(x) e^{i p y} \!\!\!
 \int\limits_{x',y'} \!\!\!
 \delta(x') e^{i p y'}
 \Big[ S_{\rm EH}[g \! = \! \delta \! + \! h^{\scriptscriptstyle \rm TT}]
 \tfrac{\overleftarrow{\delta}}{\delta h_{\mu \nu}^{\rm TT}(x')}
 \tfrac{\overleftarrow{\delta}}{\delta h_{\rho \sigma}^{\rm TT}(y')} \Big]_{h^{\! \rm TT} = 0},
 \\ \notag
 Z_{\psi}(p^{2}) ={}& \mathcal{N}_{Z_{\psi}}(p^{2}) \! \int\limits_{x,y} \! {\rm tr} \bigg( \Big[
 \Gamma_{k}[\Phi;\bar{g} = \delta] \tfrac{\overleftarrow{\delta}}{\delta \psi^{i}(x)}
 \tfrac{\overleftarrow{\delta}}{\delta \bar{\psi}^{j}(y)} \Big]_{\Phi = 0}
 \,\, \delta(x) e^{i p y}\!\!\!
 \int\limits_{x',y'} \!\!\!
 \delta(x') e^{i p y'}
 \Big[ S_{\rm mat}^{\rm kin}[\psi, \bar{\psi}; \bar{g} = \delta]
 \tfrac{\overleftarrow{\delta}}{\delta \psi^{i}(x')}
 \tfrac{\overleftarrow{\delta}}{\delta \bar{\psi}^{j}(y')} \Big]_{\substack{\psi = 0 \\ \bar{\psi} = 0}} \bigg),
 \\ \label{eq: Couplings}
 \bar{\lambda}_{2} ={}& \mathcal{N}_{\bar{\lambda}_{2}} \! \int\limits_{x,y} \! \Big[
 \Gamma_{k}[\Phi;\bar{g} = \delta] \tfrac{\overleftarrow{\delta}}{\delta h^{\mu \nu}(x)}
 \tfrac{\overleftarrow{\delta}}{\delta h^{\rho \sigma}(y)} \Big]_{\Phi = 0}
 \,\, \delta(x) \!\!\!
 \int\limits_{x',y'} \!\!\!
 \delta(x')
 \Big[ S_{\rm EH}[g \! = \! \delta \! + \! h^{\scriptscriptstyle \rm TT}]
 \tfrac{\overleftarrow{\delta}}{\delta h_{\mu \nu}^{\rm TT}(x')}
 \tfrac{\overleftarrow{\delta}}{\delta h_{\rho \sigma}^{\rm TT}(y')} \Big]_{h^{\! \rm TT} = 0},
 \\ \notag
 \bar{\lambda}_{3} ={}& \mathcal{N}_{\bar{\lambda}_{3}} \!\!\! \int\limits_{x,y,z} \!\!\! \Big[
 \Gamma_{k}[\Phi;\bar{g} = \delta] \tfrac{\overleftarrow{\delta}}{\delta h^{\mu \nu}(x)}
 \tfrac{\overleftarrow{\delta}}{\delta h^{\rho \sigma}(y)}
 \tfrac{\overleftarrow{\delta}}{\delta h^{\kappa \lambda}(z)} \Big]_{\Phi = 0}
 \,\, \delta(x) \!\!\!\!\!
 \int\limits_{x',y',z'} \!\!\!\!\!
 \delta(x')
 \Big[ S_{\rm EH}[g \! = \! \delta \! + \! h^{\scriptscriptstyle \rm TT}]
 \tfrac{\overleftarrow{\delta}}{\delta h_{\mu \nu}^{\rm TT}(x')}
 \tfrac{\overleftarrow{\delta}}{\delta h_{\rho \sigma}^{\rm TT}(y')}
 \tfrac{\overleftarrow{\delta}}{\delta h_{\kappa \lambda}^{\rm TT}(z')} \Big]_{h^{\! \rm TT} = 0},
 \\ \notag
 \bar{G}_{h} ={}& \mathcal{N}_{\bar{G}_{h}} \!\!\! \int\limits_{x,y,z} \!\!\! \Big[
 \Gamma_{k}[\Phi;\bar{g} = \delta] \tfrac{\overleftarrow{\delta}}{\delta h^{\mu \nu}(x)}
 \tfrac{\overleftarrow{\delta}}{\delta h^{\rho \sigma}(y)}
 \tfrac{\overleftarrow{\delta}}{\delta h^{\kappa \lambda}(z)} \Big]_{\Phi = 0}
 \,\, \delta(x) \, ( e^{i p_{2} y + i p_{3} z} - 1 ) 
 \\ \notag
 {}& \hspace{5.5cm} \times \!\! \int\limits_{x',y',z'} \!\!\!\!\!
 \delta(x') e^{i p_{2} y' + i p_{3} z'}
 \Big[ S_{\rm EH}[g \! = \! \delta \! + \! h^{\scriptscriptstyle \rm TT}]
 \tfrac{\overleftarrow{\delta}}{\delta h_{\mu \nu}^{\rm TT}(x')}
 \tfrac{\overleftarrow{\delta}}{\delta h_{\rho \sigma}^{\rm TT}(y')}
 \tfrac{\overleftarrow{\delta}}{\delta h_{\kappa \lambda}^{\rm TT}(z')}
 \Big]_{\!\!
 \scalebox{0.85}{$
 \substack{h^{\! \rm TT} = 0 \,\, \\ \hspace{0.3cm} \bar{\lambda} = 0 \,\, \\ \hspace{0.23cm} p^{2} = k^{2}}
 $}
 },
\end{align}
\end{widetext}
where the momenta $p_{2}$ and $p_{3}$ are evaluated at the symmetric point for three momenta with
$p_{2}^{2} = p_{3}^{2} = - 2 (p_{2} \cdot p_{3}) = k^{2}$ and
the normalizations $\mathcal{N}_{Z_{h}}$, $\mathcal{N}_{Z_{\psi}}$, $\mathcal{N}_{\bar{\lambda}_{2}}$,
$\mathcal{N}_{\bar{\lambda}_{3}}$ and $\mathcal{N}_{\bar{G}_{h}}$ are defined such, that when we plug
the $\Gamma_{k}^{(n)}$ from Eqs.~\eqref{eq: VertexGrav}, \eqref{eq: PureGravResc}, \eqref{eq: GhostResc}
and \eqref{eq: MinMattResc} into Eq.~\eqref{eq: Couplings} we get the corresponding coupling.

This projection is equivalent to the bilocal evaluation of the pure-graviton vertices, as employed, e.g.,
in \cite{Christiansen:2014raa,Christiansen:2015rva,Meibohm:2015twa,Denz:2016qks,Reichert:2018nih}.
\section{Asymptotic safety for one flavor}\label{sec:Nf1}
Phenomenologically, fermion-gravity systems with $N_{\rm f} = 24$ are of most interest,
as this is the number of Dirac fermions in the Standard Model, extended by three right-handed neutrinos.
There are indications \cite{Dona:2012am, Dona:2013qba,Meibohm:2015twa,Eichhorn:2016vvy,Alkofer:2018fxj,Alkofer:2018baq}
that such a fermion-gravity system with $N_{\mathrm{f}}>1$ features an asymptotically safe fixed point that is
continuously connected to the pure-gravity one.
We explore this hypothesis further, and therefore start by exploring a small deformation of the pure-gravity universality
class by $N_{\rm f} = 1$ fermions.

In this section, we aim at answering three key questions:
\begin{enumerate}
 \item Is there a fixed point in the fermion-gravity system that is robust under extensions of the truncation and
 changes of the projection scheme?
 \item Is the nonminimal coupling nonzero at the fixed point, and how large is its ``backreaction" onto the minimally
 coupled system?
 \item Do the avatars of the Newton coupling exhibit effective universality at this fixed point?
\end{enumerate}
\subsection{A fermion-gravity fixed point and tests of its robustness}	
\begin{table*}[!t]
 \begin{tabular}{c|c|c|c|c|c|c|c|c|c|c|c|c|c|c|c|c|c}
  Trunc & $\sigma^{*}$ & $G_{\psi}^{*}$ & $G_{h}^{*}$ & $\lambda_{2}^{*}$ & $\lambda_{3}^{*}$ & $\theta_{1}$
   & $\theta_{2/3}$ & $\theta_{4}$ & $\theta_{5}$ & $\eta_{h}(0)$ & $\eta_{h}(k^{2})$ & $\eta_{\psi}(0)$
   & $\eta_{\psi}(k^{2})$ & $A_{G_{h}}$ & $A_{{\rm f},G_{h}}$ & $A_{G_{\psi}}$ & $A_{{\rm f}, G_{\psi}}$
  \\ \hline \hline
  \vphantom{$\big(^{A}$}
  \scalebox{0.9}{$n\! = \! 1$, $p^{2} \! = \! k^{2}$}
   & $-$ & $0.79$ & $0.78$ & $0.26$ & $0.080$ & $2.8$
   & $2.6 \pm 4.0 \mathrm{i}$ & $-7.5$ & $-$ & $0.87$ & $0.12$ & $-1.4$
   & $-1.2$ & $-2.7$ & $0.023$ & $-2.6$ & $0.017$
  \\ \hline
  \vphantom{$\big(^{A}$}
  \scalebox{0.9}{$\!\! n\! = \!2$, $p^{2} \! = \! \tfrac{k^{2}}{2}, k^{2} \!$}
   & $-$ & $0.90$ & $0.75$ & $0.26$ & $0.083$ & $3.4$
   & $2.8 \pm 4.2 \mathrm{i}$ & $-7.4$ & $-$ & $0.86$ & $0.12$ & $-1.7$
   & $-1.5$ & $-2.8$ & $0.023$ & $-1.9$ & $0.033$
  \\ \hline
  \vphantom{$\big(^{A}$}
  \scalebox{0.9}{$\!\! n\! = \!2$, $p^{2} \! = \! \tfrac{k^{2}}{2}, k^{2} \!$}
   & $-0.063$ & $0.93$ & $0.75$ & $0.26$ & $0.081$ & $3.7$
   & $2.8 \pm 4.2 \mathrm{i}$ & $-7.7$ & $-2.1$ & $0.83$ & $0.08$ & $-1.6$
   & $-1.4$ & $-2.7$ & $0.005$ & $-1.6$ & $-0.002$
 \end{tabular}
 \caption{
 Fixed-point values as well as coefficients of the beta functions in the three schemes.
 The classification of different truncations refers to the order in the $n$-sample point projection.
 Here the $\theta_{i}$ are the critical exponents, i.e., the
 eigenvalues of the stability matrix multiplied by an additional negative sign.
 For the definition of the $A$'s see Eq.~\eqref{eq:betaA}.
 }
 \label{tab: BigTable}
\end{table*}
The $n\!\!=\!\!1$ truncation, i.e., $\sigma = 0$, exhibits an interacting fixed point providing further evidence
for the existence of the Reuter fixed point in gravity-fermion systems.
The fixed-point results in the first two lines of Tab.~\ref{tab: BigTable} are well-compatible with those of
\cite{Eichhorn:2018ydy}, calculated for a different value of the gauge-parameter $\beta$, for a similar system.
The Reuter fixed point is characterized by three relevant and one irrelevant direction in this truncation,
cf.~Tab.~\ref{tab: BigTable}.
Due to the large overlap of the corresponding eigendirection with $\lambda_{3}$, the negative critical exponent
can be associated with this direction in theory space.
Thus, while $\lambda_{2}$ remains relevant, $\lambda_{3}$ is shifted into irrelevance at the UV fixed point.
The most relevant direction, corresponding to the critical exponent $\theta_{1}$ in \autoref{tab: BigTable}
has largest overlap with $G_{\psi}$, while the remaining two directions form a complex pair.
We highlight that only a subset of these critical exponents is physical, namely those in the
diffeomorphism invariant theory space.
Here, we are exploring a larger theory space that accounts for symmetry-breaking through the regulator
and gauge-fixing terms, and therefore includes different avatars of the Newton coupling, as well as
the two couplings $\lambda_{2}$, $\lambda_{3}$.
Of the four critical exponents, only two are therefore physical.
The observation that $\theta_{1}$ is essentially associated with $G_{\psi}$ can be interpreted as a hint
that diffeomorphism-symmetry restoration in the flow towards the physical point $k = 0$ is possible, as a
relevant coupling can be chosen arbitrarily in the IR.

The robustness of these results can be tested by changing the projection scheme for $G_{\psi}$ from
$n\!\!=\!\!1$ to $n\!\!=\!\!2$ for the number of sample-points, while staying within the same truncation.
As shown in the second line of Tab.~\ref{tab: BigTable} this leads to a change of about $14 \%$ for the fixed-point
value of $G_{\psi}$, but more importantly a change of about $21 \%$ at the level of the universal%
\footnote{%
With universality at the level of critical exponents we refer to the insensitivity of critical exponents
to unphysical details like gauge, regulator and scheme.
We refer to universality in the larger theory space, accordingly a universal critical exponent is not automatically
associated to physics.
}
critical exponent $\theta_{1}$.
This indicates the necessity of extensions of the truncation.
We stress that the change of the other couplings and critical exponents stays small, as the feedback of $G_{\psi}$ into
the other beta functions is with a very small prefactor.
The relative contribution of $G_{\psi}$ versus that of $G_{h}$ in $\beta_{G_{h}}$ is for instance approximately given by
$A_{{\rm f}, G_{h}}/A_{G_{h}}$.
Here, $A_{G_{i}}$ is the $N_{\rm f}$-independent and $A_{{\rm f}, G_{i}}$ the prefactor for the $N_{\rm f}$-dependent
quadratic coefficient of $\beta_{G_{i}}$ once both avatars are equated,
\begin{equation} \label{eq:betaA}
 \beta_{G_{i}} \big|_{G_{h} =G_{\psi} = G}
 = 2 G + G^{2} (A_{G_{i}} + N_{\mathrm{f}} A_{\mathrm{f}, G_{i}}) + \mathcal{O}(G^{3}) \, .
\end{equation}
Accordingly, a change of the projection scheme for $G_{\psi}$, accompanied by an unchanged projection for the other
couplings, is not expected to result in significant changes in any values except for $G_{\psi}^{*}$ and $\theta_{1}$.
Hence, the small relative changes in $G_{h}^{\ast}, \lambda_{2}^{\ast}, \lambda_{3}^{\ast}$ and $\theta_{2,3,4}$
should not be taken as an indication of quantitative convergence  in the system.

A further aspect providing information on the quality of our truncation
is the momentum dependence of the vertex flow.
In Sec.~\ref{sec:betas}, we derived a formal expression Eq.~\eqref{eq: vertflow} for the beta function of the fully momentum
dependent vertex function $V\big(\tfrac{p^{2}}{k^2}\big)$.
Evaluated at the full fixed point $V^{\ast}$ in \emph{untruncated} theory space,
the right-hand side of Eq.~\eqref{eq: vertflow} would vanish for all momenta
 -- assuming that such a fixed point indeed exists.
In truncations this is not the case.
Within a given truncation scheme, we can use the truncated beta functions, cf.~Eqs.~\eqref{eq: betaAnsatz},
to find a  fixed point $V_{\rm trunc}^{\ast}$
of the truncated RG flow projected onto the subspace defined by the truncation.
However, this truncated fixed point will not satisfy the fixed-point equation for all momenta,
i.e., $\beta_{V}\big( \tfrac{p^{2}}{k^{2}} \big)|_{V = V_{\rm trunc}^{\ast}} \neq 0$.
This is a consequence of the fact that the RG flow in truncations is not closed.
Terms beyond the truncation are generated.
Accordingly, the RG flow features additional, higher-order momentum dependence than what is captured by the truncation.
Thus we introduce the quantity $\mathcal{V}$, which allows us to estimate the deviation
of the truncated fixed point $V_{\rm trunc}^{\ast}$ from the full fixed point $V^{\ast}$,
\begin{equation} \label{eq: MomDeps}
 \mathcal{V} \big( \tfrac{p^{2}}{k^2} \big)
 = \beta_V \big( \tfrac{p^{2}}{k^2} \big)|_{V=V^{\ast}_{\rm trunc}} + V_{\rm trunc}^{\ast} \big( \tfrac{p^{2}}{k^2} \big).
\end{equation}
Specifically, the idea is simply that if we can satisfy
$\beta_V \big( \tfrac{p^{2}}{k^2} \big)|_{V=V^{\ast}_{\rm trunc}} \approx 0$ for all momenta,
then we expect that the fixed point in our truncation comes close to the fixed point in untruncated theory space.
Away from the fixed point $\mathcal{V}$ is an
auxiliary quantity and has no direct physical meaning.
For a fixed point in untruncated theory space, the function $\beta_{V}\big(\tfrac{p^{2}}{k^2}\big)|_{V=V^{\ast}}$
vanishes, such that $\mathcal{V}=V^{\ast}$.
This is a general self-consistency equation for the fixed point of the system at any value of $p^{2}$.
Our $\mathcal{V}$ is similar to what has been investigated in
\cite{Denz:2016qks, Eichhorn:2018akn}.%
\footnote{
There, the auxiliary quantity was defined as
$\mathcal{V}_{\rm Lit} \big(p^{2}/k^{2} \big)
= - \frac{2 \beta_{V}(p^{2}/k^{2}) }{2 + \eta_{h}(p^{2}) + 2\eta_{\psi}(p^{2})}\big|_{V=V_{\rm trunc}^{\ast}}
+ V_{\rm trunc}^{\ast}(p^{2}/k^{2})$,
which only differs from $\mathcal{V}$ by a normalization in front of $\beta_V$.
We chose the definition \eqref{eq: MomDeps} to avoid artificial poles that might arise due to the denominator in
$\mathcal{V}_{\rm Lit}$.
}\\
To further explain the meaning of Eq.~\eqref{eq: MomDeps}, we now specialize to our
truncation where $V_{\rm trunc}\big(\tfrac{p^{2}}{k^2}\big)$ is a polynomial in $\tfrac{p^{2}}{k^2}$ only up to
first order, cf.~Eq.~\eqref{eq: Ansatz}, while ${\rm Flow}^{(3)}_{\psi}$ clearly contains higher powers.
Therefore, $\mathcal{V}$ can at best be approximately equal to $V_{\rm trunc}^{\ast}$.
Thus, the difference of our choice for $V_{\rm trunc}^{\ast}\big(\tfrac{p^{2}}{k^2}\big)$ given
in Eq.~\eqref{eq: Ansatz} from $\mathcal{V}\big(\tfrac{p^{2}}{k^2}\big)$ shows how well
$V_{\rm trunc}^{\ast}\big(\tfrac{p^{2}}{k^2}\big)$ approximates the momentum dependence of
the full fixed point $V^{\ast}\big( \tfrac{p^{2}}{k^{2}} \big)$.

\begin{figure}[!t]
 \centering
 \includegraphics[width=1\linewidth]{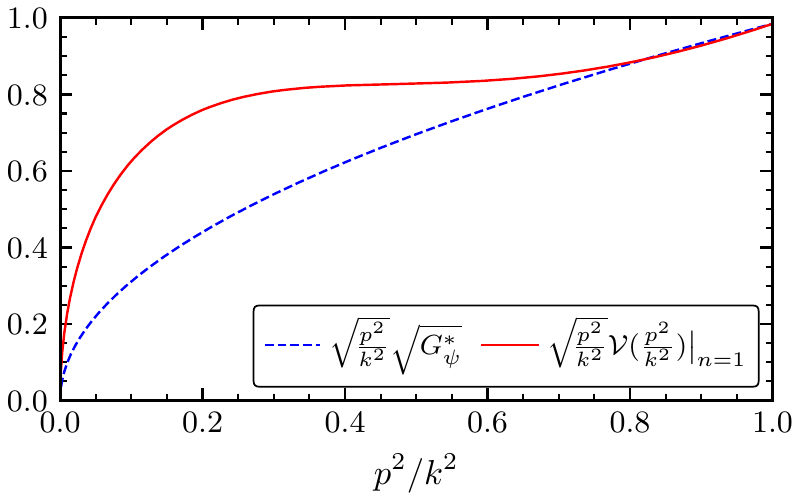}
 \caption{
 Momentum dependence of the graviton-fermion three-point vertex, evaluated with $n = 1$ sample-points,
 at the corresponding fixed point.
 The dashed blue line corresponds to the truncated vertex $V_{\rm trunc}^{\ast}$,
 which is equal to $\sqrt{G_{\psi}^{\ast}}$ in the $n=1$ scheme.
 Note that $V_{\rm trunc}$ always enters the flow equation with an additional factor $\tfrac{p}{k}$.
 Thus we plot $\tfrac{p}{k} V_{\rm trunc}^{\ast}$.
 The solid red line corresponds to the full momentum dependence of the auxiliary quantity
 $\tfrac{p}{k} \mathcal{V} = \tfrac{p}{k} (\beta_{V}|_{V = V_{\rm trunc}^{\ast}} + V_{\rm trunc}^{\ast} )$,
 cf.~Eq.~\eqref{eq: MomDeps} that is generated from within our truncation, but goes beyond the momentum-dependence
 of the term in our truncation.
 The difference between the two lines indicates the need for an extension of the truncation,
 as both would agree if evaluated at an untruncated fixed point $V^{\ast}$
 due to the vanishing of $\beta_{V}|_{V = V^{\ast}}$.}
\label{fig: Biloc}
\end{figure}

\begin{figure}[!t]
 \centering
 \includegraphics[width=1\linewidth]{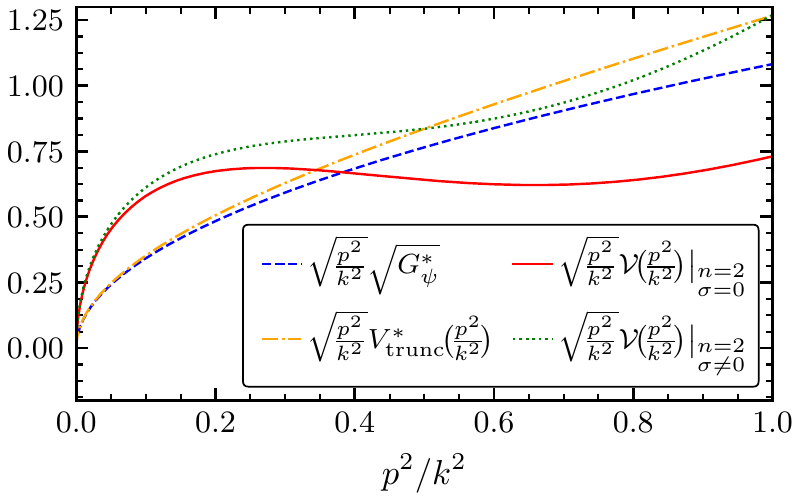}
 \caption{
 Momentum dependence of the graviton-fermion three-point vertex evaluated at $n\!\!=\!\!2$ sample-points,
 at the corresponding fixed point.
 The dashed blue and the dot-dashed orange lines correspond to the ansatz for the vertex $\tfrac{p}{k} V_{\rm trunc}^{\ast}$
 \eqref{eq: Ansatz}, setting $\sigma = 0$ for the blue dashed line and including $\sigma^*$ for the orange dot-dashed one.
 The solid red and dotted green lines corresponds to the full momentum dependence given by
 $\tfrac{p}{k}\mathcal{V}=\tfrac{p}{k}\left(\beta_V|_{V = V_{\rm trunc}^{\ast}} + V_{\rm trunc}^{\ast}\right)$
 in Eq.~\eqref{eq: MomDeps}.
 }\label{fig: Triloc}
\end{figure}

In summary, Eq.~\eqref{eq: MomDeps} can be read as follows: $\mathcal{V}\big(\tfrac{p^{2}}{k^2}\big)$
captures the momentum-dependence as generated by the flow of the fermion-gravity vertex,
i.e., by the diagrams in Fig.~\ref{fig:flowdiags}.
In untruncated theory space, the full momentum dependence would be captured by $V^{\ast}\big( \tfrac{p^{2}}{k^{2}} \big)$,
leading to $\beta_V\big(\tfrac{p^{2}}{k^2}\big)|_{V=V^{\ast}} =0$.
In truncations, the ansatz $V_{\rm trunc}^{\ast}\big(\tfrac{p^{2}}{k^2}\big)$ differs from the momentum dependence
$\mathcal{V}\big(\tfrac{p^{2}}{k^2}\big)$ generated by the flow, such that
$\beta_V\big(\tfrac{p^{2}}{k^2}\big)|_{V = V_{\rm trunc}^{\ast}}$ vanishes only at selected values of $\tfrac{p^{2}}{k^2}$.
Accordingly, the comparison of $\mathcal{V}\big(\tfrac{p^{2}}{k^2}\big)$ and
$V_{\rm trunc}^{\ast}\big(\tfrac{p^{2}}{k^2}\big)$ is well suited to check whether higher-order momentum dependences
beyond the truncation are generated by the flow equation.
Moreover, the magnitude of higher-order momentum dependences can be estimated.
Finally, the flow can of course show a different momentum dependence at large and small momenta;
e.g., being well-approximated by a simple low-order polynomial at $\tfrac{p^{2}}{k^2}\approx 1$,
and exhibiting a more intricate momentum-dependence for $\tfrac{p^{2}}{k^2}\approx 0$.
Comparing $\mathcal{V}\big(\tfrac{p^{2}}{k^2}\big)$ and $V_{\rm trunc}^{\ast}\big(\tfrac{p^{2}}{k^2}\big)$ at all values of
$\tfrac{p^{2}}{k^2}$ provides information on such cases.
This provides another guiding principle on how to extend the truncation:
The momentum-dependence of the flow, evaluated at the fixed-point values in the $n\!\!=\!\!1$ and $n\!\!=\!\!2$-sample-point
projection with $\sigma=0$ both indicate the presence of a $\tfrac{p^{3}}{k^{3}}$ term in the flow of the
graviton-fermion vertex, cf.~Fig.~\ref{fig: Biloc} and Fig.~\ref{fig: Triloc}.
There, we plot the comparison of both sides of Eq.~\eqref{eq: MomDeps}, multiplied by $\tfrac{p}{k}$.
This is motivated by the classical structure of the graviton-fermion vertex, which takes the form
$\tfrac{p}{k} V_{\rm trunc}\big(\tfrac{p^{2}}{k^2}\big)$.
The fact that Eq.~\eqref{eq: Ansatz} does not carry this factor of $\tfrac{p}{k}$ is due to the specific choice
of the projector Eq.~\eqref{eq: Proj3}.
Therefore, Fig.~\ref{fig: Biloc} and Fig.~\ref{fig: Triloc} show the vertex $V_{\rm trunc}^{\ast}\big(\tfrac{p^{2}}{k^2}\big)$
weighted by how it contributes in the diagrams.
The presence of a $p^3$ contribution in Fig.~\ref{fig: Biloc} and Fig.~\ref{fig: Triloc} motivates our extension
of the truncation by $\sigma$ that we explore below.
This extension allows us to feed the higher momentum dependence, which is already seen in the projections with $\sigma=0$,
back into the diagrams and therefore consider this information.
\subsection{Generation and backreaction of \texorpdfstring{$\sigma$}{sigma}}\label{subsec:sigma}
Motivated by the observation that higher-momentum dependence is clearly generated by the flow equation,
we make the next step in extensions of the truncation following canonical power-counting.
In the present truncation, the next-to-leading-order coupling that contributes to the graviton-fermion vertex is the
nonminimal coupling $\sigma$.
The interaction itself is dimension-6, accordingly the canonical dimension of the coupling is $-2$.
Symmetry-arguments elaborated on in \cite{Eichhorn:2017eht} indicate that it should not be possible to realize
$\sigma^{\ast} = 0$, once $G_{\psi}^{\ast} \neq 0$.
Here we will check whether this is indeed the case in the corresponding truncation and will further explore the backreaction
of $\sigma$ on the system of couplings in the smaller truncation.
For the latter, we also focus on the momentum-dependence of the flow, to find whether it is captured more adequately once
the next-to-leading term beyond $G_{\psi}$ is included.

At the fixed point for the nonminimally coupled fermion-gravity system in the $n\!\!=\!\!2$ projection,
the flow of the nonminimal coupling $\sigma$ is given by:
\begin{align} \label{eq:betasigma}
 \beta_{\sigma} \bigg|_{\lambda_{2}^{*}, \lambda_{3}^{*}, G_{h}^{*}, G_{\psi}^{*}}
 = 0.13 + 2.0 \sigma - 1.9 \sigma^{2} + 1.6 \sigma^{3} + \mathcal{O}(\sigma^4) \, .
\end{align}
The existence of the $\sigma$-independent term confirms that the nonminimal coupling is indeed induced at the UV fixed-point
for gravity.
The same conclusion can be drawn in the derivative expansion, as indicated by various $\sigma$-independent terms in
Eq.~\eqref{eq:appbetasigexp} and Eq.~\eqref{eq:appbetasig}.
As a consequence of such terms, $\sigma^{*} = 0$ is not a solution of the flow equation and the nonminimal coupling cannot
be consistently set to zero.
The fixed-point value of $\sigma$ is indeed finite, cf.~Tab.~\ref{tab: BigTable}.
Further, the inclusion of $\sigma$ gives rise to an additional irrelevant direction, cf.~Tab.~\ref{tab: BigTable}.
Due to the negative canonical mass dimension, this is expected.
It provides yet another indication that the critical hypersurface of the Reuter fixed point is finite dimensional.
The shift between the canonical dimension $d_{\bar{\sigma}}=-2$ and the critical exponent is significantly less than one.
This provides a further check for the hypothesis that the critical hypersurface of the Reuter fixed point has a
low dimensionality, with no more relevant directions than canonically relevant and marginal couplings.
Further, the difference $|\theta_{5} - d_{\bar{\sigma}}| \ll 1$ is in line with the potential near-perturbative
nature of the fixed point, cf.~Sec.~\ref{subsec:effuni} below.

A key question is the size of the ``backreaction'' of $\sigma$ on the minimally coupled system.
This provides an indication about the state of convergence of the truncation, at least in this particular direction
in theory space and thereby it provides guidance about the setup of future truncations.
In \cite{Eichhorn:2017eht}, it was shown that in the part of the gravitational coupling-space where fixed-point values
lie at small numbers of matter fields, the ``backreaction'' of induced matter self-interactions onto the remaining system
is subleading compared to the direct gravity contribution.
Here, we make a similar observation for the nonminimal matter-gravity coupling.
Due to the small fixed-point value for the nonminimal coupling $\sigma^{*} = - 0.063$, the impact of $\sigma$ on
the minimally coupled system is small.
In fact, the smallness of $\sigma^{\ast}$ is crucial in view of its negative sign (see also Fig.~\ref{fig: MaxMinPosRescal}):
For a sufficiently negative fixed-point value, $\sigma$ can alter the effect of fermionic fluctuations on the Newton
couplings from screening to antiscreening, see also Eq.~\eqref{eq: GpsiDeriv} and Eq.~\eqref{eq: GhDeriv}
in App.~\ref{sec:appderbeta}.
Indeed for $N_{\mathrm{f}}=1$, the fixed-point value for $\sigma$ leads to a slightly negative $A_{\mathrm{f}, G_{\psi}}$.
Yet one should keep in mind that within the systematic error of $A_{\mathrm{f}, G_{\psi}}$, estimated, e.g.,
by the difference of $A_{\mathrm{f}, G_{\psi}}$ in the two projection schemes, cf.~first two lines
in Tab.~\ref{tab: BigTable}, $A_{\mathrm{f}, G_{\psi}}$ is compatible with being positive also in the presence of $\sigma$.
For $N_{\rm f}>1.5$, the effect of fermionic fluctuations on both avatars of the Newton coupling is screening, i.e.,
$A_{\mathrm{f}, G_{\psi}} > 0$.
Taken together, we view this as tentative evidence to consider the sign of $A_{\mathrm{f}, G_{\psi}}$ at $N_{\mathrm{f}}=1$
as an artifact of unphysical choices, such as gauge, regulator and projection scheme.

Due to the smallness of $\sigma^{\ast}$, the approximation of the momentum dependence does not differ significantly from a
straight line, cf.~Fig.~\ref{fig: Triloc}, as the $p^{2}$ contribution introduced by $\sigma$ is small at the fixed point.
However, the inclusion of $\sigma$ nevertheless leads to a better approximation of the momentum structure at
$p^{2} > 0.4 k^{2}$, compared to the $n\!\!=\!\!2$ projection where $\sigma$ was neglected, cf.~Fig.~\ref{fig: Triloc}.

To provide a more intuitive comparison of how well different truncations capture the full momentum dependence
of the vertex flow, we directly evaluate $\beta_{V}$ given by Eq.~\eqref{eq: vertflow} for the different truncations.
Evaluated at the fixed point, this expression should vanish for all values of $p^{2}$.
Thus, the deviation of the right-hand side, which we denote by $\beta_V\big|_{V = V_{\rm trunc}^{\ast}}$,
encodes the deviation of a given truncation from the full momentum structure.%
\footnote{%
Note that $\beta_V\big|_{V = V_{\rm trunc}^{\ast}}$ denotes the full $\beta_V$ \eqref{eq: vertflow}
evaluated at the truncated fixed point $V_{\rm trunc}^{\ast}$, as opposed to $\beta_V^{\rm trunc}$ \eqref{eq: betaAnsatz},
which only contains the beta functions calculated in our truncation.
}
It describes how accurately the truncation approximates the fixed-point Eq.~\eqref{eq: vertflow}
for all values of $p^{2}$.
One can interpret $\beta_V\big|_{V = V_{\rm trunc}^{\ast}}$ simply as a test, whether the momentum-dependence of the
right-hand-side of the Wetterich equation is described accurately by a constant and a $p^{2}$ term with $k$-dependent but
$p$-independent couplings.
If our truncation was exact, such that upon input of $G_{\psi}$  and $\sigma$ no additional terms were generated on the
right-hand-side, then $\beta_{V}\big|_{V = V_{\rm trunc}}$ would vanish for all momenta once the fixed-point values for
the couplings are inserted.
Compared to our previous tests shown in Fig.~\ref{fig: Biloc} and Fig.~\ref{fig: Triloc}, this allows for a more direct
comparison of different truncations.
As a drawback, the information on which powers of $\tfrac{p^{2}}{k^2}$ are induced in each truncation is somewhat
less obvious.

\begin{figure}[!t]
  \centering
  \includegraphics[width=1\linewidth]{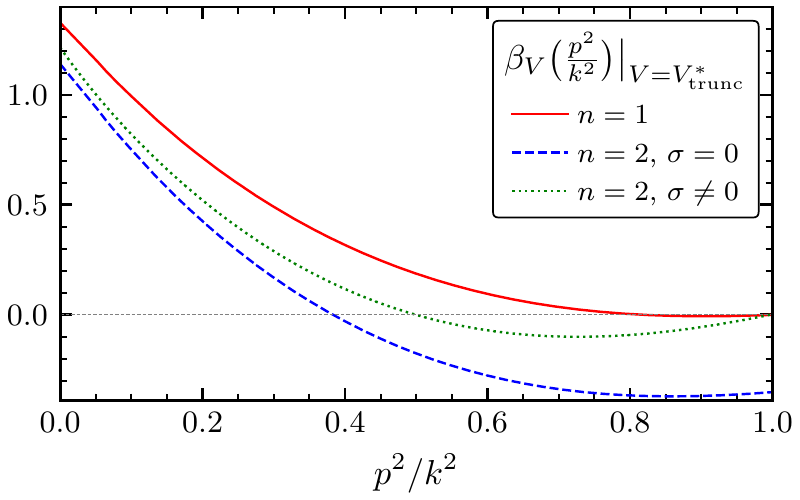}
  \caption{
  As a key result of this section, we compare all different approximations of the momentum dependent flow.
  Here $\beta_{V}\big|_{V = V_{\rm trunc}^{\ast}}$ refers to the evaluation of the right-hand side of
  Eq.~\eqref{eq: vertflow} for a given truncation at the corresponding fixed point $V_{\rm trunc}^{\ast}$.
  The deviation of this expression from zero encodes the accuracy of the projection.
  We highlight that the $n\!\!=\!\!1$ scheme and $n\!\!=\!\!2$ scheme including $\sigma$ capture the full
  momentum-dependence at $p^2=k^2$ by construction.
  Additionally, the $n\!\!=\!\!2$ scheme including $\sigma$ leads to a lower value of $\beta_V|_{V = V_{\rm trunc}^{\ast}}$,
  i.e., a better approximation of the full flow, also at lower values of the momentum.
  } \label{fig: MomComp}
\end{figure}

Fig.~\ref{fig: MomComp} shows this quantity evaluated at the corresponding fixed point in all different truncations.
We caution that the three curves in Fig.~\ref{fig: MomComp} have a systematic error due to our truncation.
Therefore not all differences between the curves are significant.
\begin{figure}[!t]
  \centering
  \includegraphics[width=1\linewidth]{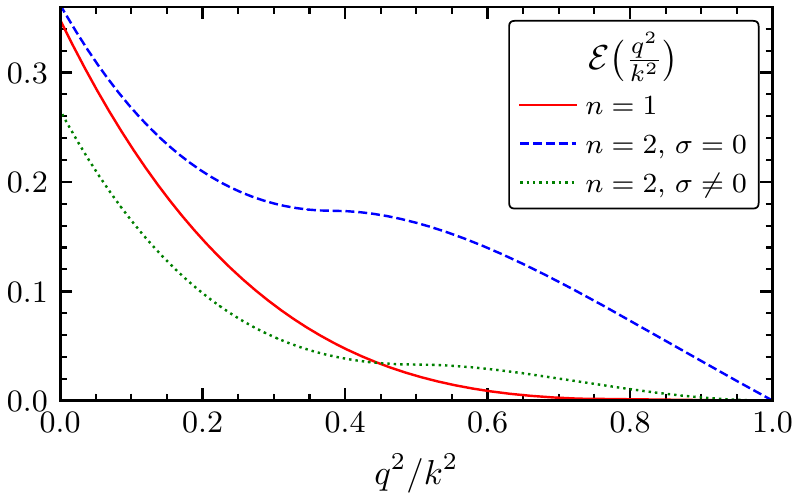}
  \caption{
  Integrated deviation in the momentum dependence in different truncations as defined in Eq.~\eqref{eq:dev}.
  We integrate from $p=k$ down to $p=q$.
  Here $\mathcal{E}$ refers to the evaluation of Eq.~\eqref{eq:dev} for a given truncation at the
  fixed point in the particular truncation.
  The deviation of this expression from zero encodes the accuracy of the projection.
  }\label{fig: IntMomComp}
\end{figure}

Two aspects of the curves are of particular interest to us.
Firstly, momenta $p^{2} \approx k^{2}$ are expected to provide the main contribution to the flow.
Accordingly, it is desirable for an approximation to capture the momentum dependence at $p^{2} \approx k^{2}$ accurately.
This happens if $\beta_V\big|_{V = V_{\rm trunc}^{\ast}} \approx 0$ for these momenta.
Secondly, all momenta (except $p^{2} = 0$) contribute to the flow, even if momenta with $p^{2} \ll k^{2}$ contribute less.
Accordingly it is desirable to minimize the integrated deviation of $\beta_V\big|_{V = V_{\rm trunc}^{\ast}}$ from zero.
Fig.~\ref{fig: MomComp} serves as a summary of our results for the momentum-dependence of the system:
While the momentum-dependence of the flow is well-captured by the $n\!\!=\!\!1$ scheme and the $n\!\!=\!\!2$ scheme
including $\sigma$, the latter highlights the necessity for further extensions of the truncation, i.e.,
inclusion of higher-order operators.
Further, the $n\!\!=\!\!2$ scheme including $\sigma$ captures the momentum-dependence slightly better at low values of $p^2$.
Nevertheless, the performance of the $n\!\!=\!\!1$ scheme and the $n\!\!=\!\!2$ scheme including $\sigma$ are comparable
at $N_{\rm f}=1$.

To provide a quantitative characterization, we define
\begin{equation} \label{eq:dev}
 \mathcal{E}\left(\frac{q^2}{k^2}\right) =\int_{\frac{q^2}{k^2}}^{1} {\rm d} y \, \beta_V (y)|_{V = V_{\rm trunc}^{\ast}}.
\end{equation}
For a truncation that captures the full momentum dependence, $\mathcal{E}=0$.
As we are mostly interested in capturing the momentum dependence around $p^2 \approx k^2$ correctly,
we integrate $\beta_V$ from a lower boundary $\tfrac{q^2}{k^2}$ up to $1$, cf.~Fig.~\ref{fig: IntMomComp}.
We find that the $n=2$ projection with $\sigma=0$ performs worst.
The $n\!\!=\!\!1$ projection matches the flow most closely down to $\tfrac{q^2}{k^2} \approx 0.5$.
In this region, the approximation with $n\!\!=\!\!2$ sample points only yields a small integrated deviation of order $0.04$.
The total integrated deviation is smallest for the $n\!\!=\!\!2$ projection with the inclusion of $\sigma$,
indicating that it captures the full momentum dependence best. 

As a tentative conclusion of our analysis, we emphasize that the inclusion of $\sigma$ leads to a setting which captures
the momentum-dependence of the full flow with a reasonable accuracy.
Nevertheless, both Fig.~\ref{fig: Triloc} and Fig.~\ref{fig: MomComp} suggest that full accuracy requires further
extensions of the truncation to include higher-order momentum dependencies.
\subsection{Effective universality}\label{subsec:effuni}
In a theory space where a gauge symmetry is broken, different avatars of this gauge coupling no longer agree.
If the symmetry-breaking is a consequence of gauge fixing and the regulator, then modified Slavnov-Taylor identities
select a hypersurface in this theory space.
This hypersurface  does not coincide with the symmetric theory space.
It reduces to that of the standard Slavnov-Taylor identities at $k = 0$,
where the regulator vanishes.
The initial condition for the flow at $k\rightarrow \infty$ should contain symmetry-breaking terms such that
the symmetry-breaking introduced by the regulator during the flow can be compensated for by the initial condition,
see the pedagogical introduction in \cite{Gies:2006wv}.
Accordingly, the fixed point might exhibit a difference between distinct avatars of the gauge coupling.
Additionally, there is a modified shift Ward-identity, see, e.g., \cite{Manrique:2009uh} as well as
\cite{Reuter:1997gx,Litim:2002ce,Pawlowski:2005xe} for gauge theories, encoding the difference between correlation
functions of the background field and the fluctuation field, which has been explored, e.g.,
in \cite{Morris:2016spn,Percacci:2016arh,Labus:2016lkh,Ohta:2017dsq,Eichhorn:2018akn}.
Effective universality was defined in \cite{Eichhorn:2018akn, Eichhorn:2018ydy} as a near-agreement of different avatars
of the Newton coupling.
In the case of Yang-Mills theories in four dimensions with a marginal gauge coupling, effective universality holds in
the perturbative regime as a consequence of two-loop universality:
Up to two loops, the avatars of the gauge coupling agree exactly.
Within perturbation theory, the higher-loop terms are subleading.
Accordingly, effective universality follows due to the marginal nature of the gauge coupling.
In quantum gravity, the case is more subtle.
As the deviation from universality is a consequence of quantum effects (the different avatars agree trivially in
a classical theory), the perturbative regime, where quantum effects are small, is expected to exhibit effective universality.

Effective universality is not a feature that a viable asymptotically safe fixed point necessarily must exhibit.
Instead, it can be viewed as a criterion that distinguishes nonperturbative from near-perturbative fixed points.
Other hints at a near-perturbative nature of the Reuter fixed point consist in the near-canonical scaling spectrum of
higher-order curvature operators
\cite{Falls:2013bv,Falls:2014tra,Falls:2017lst,Falls:2018ylp} which is also exhibited by $\sigma$ in our system,
and the possibility to uncover the fixed point within one-loop perturbation
theory \cite{Niedermaier:2009zz,Niedermaier:2010zz}.
Further, the Reuter fixed point might be connected continuously to the perturbative asymptotically safe fixed point
seen within the epsilon expansion around $d = 2$ dimensions \cite{Weinberg:1980gg,Kawai:1992np}.
Indications for this have been found, e.g., in \cite{Codello:2008vh,Nink:2012vd,Nink:2015lmq,Falls:2015cta},
however, see also \cite{Litim:2003vp,Biemans:2016rvp}.

Moreover, such a feature appears desirable from a phenomenological point of view:
According to the findings in \cite{Eichhorn:2017ylw,Eichhorn:2017lry,Eichhorn:2018whv}, a near-perturbative gravitational
fixed point could induce an asymptotically safe UV completion of the Standard Model, which matches onto the perturbative
RG flow of the Standard Model from the Planck scale to the IR.

To investigate whether the effective universality is realized in this system we expand the beta functions for both
avatars of the Newton coupling at the corresponding fixed-point values for $\lambda_{2}$ and $\lambda_{3}$,
cf.~Eq.~\eqref{eq:betaA}.
The mechanism behind asymptotic safety in gravity is a cancellation of canonical scaling with quantum scaling.
The quantum term must have an antiscreening nature to generate a viable fixed point at $G^{\ast} > 0$.
For a fixed point which can be traced back to the free fixed point as $d \rightarrow 2$, the quantum effects are
captured qualitatively by the leading term in the expansion in $G$.
Therefore we consider the two quadratic coefficients $A_{G_{i}}$ and $A_{{\rm f}, G_{i}}$ as encoding key physics
of asymptotic safety.
The comparison of these coefficients for different avatars of the Newton coupling also allows to deduce
whether effective universality is realized.

We discover a quantitative agreement of the two avatars in the $n\!\!=\!\!1$ projection.
This holds for the $N_{\mathrm{f}}$-independent part, with $A_{G_{\psi}}/A_{G_{h}} \approx 0.96$ as well for the
$N_{\mathrm{f}}$-dependent part with $A_{\mathrm{f},G_{\psi}}/A_{\mathrm{f},G_{h}} \approx 0.73$.
The close agreement of these coefficients is reflected in the good agreement of $G_{h}^{\ast}$ and $G_{\psi}^{\ast}$.

A measure for the deviation from effective universality has been introduced in \cite{Eichhorn:2018akn}:
\begin{align} \label{eq: epsEffUniv}
 \varepsilon(G, \mu, \lambda_{3})
 = \left|
 \frac{ \Delta \beta_{G_{h}} - \Delta \beta_{G_{\psi}} }{ \Delta \beta_{G_{h}} + \Delta \beta_{G_{\psi}} }
 \right|_{G_{h} = G_{\psi} = G} \, ,
\end{align}
where $\Delta \beta_{G_{i}} = \beta_{G_{i}} - 2 G_{i}$ is the anomalous part of $\beta_{G_{i}}$.
Expressed in this measure, $\varepsilon \approx 0.01$ at the interacting fixed point of the $n\!\!=\!\!1$ truncation.
This value indicates an almost exact agreement of the beta functions of both avatars of the Newton coupling.
The systematic error of our truncation has been estimated in \cite{Eichhorn:2018ydy} to result in an uncertainty
$\delta \varepsilon \approx 0.2$.
Within this estimate for the systematic error, the fixed point is compatible with exact universality according to
the measure in Eq.~\eqref{eq: epsEffUniv}.
This indicates the near-perturbative nature of the interacting fixed point in this truncation, as pointed out in
\cite{Eichhorn:2018ydy}.

Evaluating this measure of effective universality in the $n\!\!=\!\!2$ projection without the inclusion of $\sigma$
yields a value of $\varepsilon \approx 0.2$.
While the value is larger than that in the $n\!\!=\!\!1$ projection, it is still compatible with exact universality within
the estimate for the systematic uncertainty of $\varepsilon$.
The discovery of effective universality in this system accordingly appears to be quite robust.

At a first glance, the presence of $\sigma$ at the fixed-point could look like a potential source of deviation from
effective universality.
After all, $\sigma$ couples differently into $\beta_{G_{h}}$ than it does into $\beta_{G_{\psi}}$.
Yet, we caution that this conclusion might be premature:
If the Reuter fixed point features effective universality, this is a consequence of many structurally different
contributions in the beta functions.
At the fixed point, nontrivial cancellations occur which result in effective universality.
In fact, this is the case in the truncation explored in \cite{Eichhorn:2018akn,Eichhorn:2018ydy}:
Although $\lambda_{2}$ and $\lambda_{3}$ couple differently into $\beta_{G_{h}}$ and $\beta_{G_{\psi}}$,
effective universality is realized.
Hence, a priori one cannot infer whether or not the inclusion of $\sigma$ will spoil effective universality,
as a dynamical adjustment of fixed-point values can lead to effective universality at one point in theory space
even though the impact of $\sigma$ on $\beta_{G_{h}}$ and $\beta_{G_{\psi}}$ is structurally different.
Therefore we now explicitly check this question.
Including $\sigma$, we find $\varepsilon \approx 0.26$.
Within our estimate for the systematic uncertainty of $\varepsilon$, this appears to be compatible with effective
universality defined as a near-agreement of the fixed-point values of the avatars, even though it appears to be just
incompatible with an exact agreement.

In summary, both truncations ($\sigma =0$ and $\sigma\neq 0$), with the $n\!\!=\!\!1$ and $n\!\!=\!\!2$
projection schemes that we have explored indicate the possibility of effective universality at the Reuter fixed point,
hinting at a potentially near-perturbative nature of the fixed point.
\section{Structural aspects of the weak-gravity bound} \label{sec:wgb}
In this section, we broaden our view away from the fixed point in the above truncation, and instead analyze
$\beta_{\sigma}$ with all other couplings treated as external parameters.
Varying these allows us to explore the behavior of the system away from the results in one specific truncation.
For this section, we assume effective universality, i.e., $G_{\psi}= G_h=G$, motivated by our results in the previous section.
Further, we stress that our analysis for the critical exponent from $\beta_{\sigma}$ can be translated to the larger
system in which all couplings are dynamical, if the stability matrix is approximately block diagonal.
Our results in the previous section highlight that this is the case, at least at the fixed point explored there.

In \cite{Eichhorn:2016esv, Christiansen:2017gtg, Eichhorn:2017eht}, the weak-gravity bound for
asymptotic safety was introduced.
It is based on the observation that strong metric fluctuations can lead to the loss of a predictive fixed point in
matter interactions.
Specifically, these are such couplings that cannot be set to zero in the presence of asymptotically safe gravity.
Due to the symmetry-structures in the matter sector, these couplings are all canonically irrelevant.
For a subset of those, analyzed in truncated flows, strong metric fluctuations lead to a loss of the shifted
Gaussian fixed point (sGFP) at a fixed-point collision.
The maximum strength of gravitational fluctuations that is still compatible with a real shifted Gaussian fixed point
leads to a bound on the gravitational fixed-point values, the weak-gravity bound.
We stress that even in the region beyond the weak-gravity bound, the beta functions might allow for other zeros to exist.
These additional zeros need not correspond to actual fixed points and could be truncation artifacts.
More importantly, these zeroes are typically associated with a critical exponent that deviates rather significantly
from a canonical power-counting, invalidating the truncation scheme that is typically used.
In particular, the couplings in question are all canonically irrelevant in $d = 4$, but might be relevant at the additional
zero of the beta function.
This would imply the existence of an additional free parameter, corresponding to a reduced predictivity.
In particular, when it comes to matter couplings, there are no experimental indications that such free parameters
beyond the couplings of the Standard Model exist in nature.
Therefore, the region beyond the weak-gravity bound, where the sGFP ceases to exist, is not strictly excluded as a viable
region for the Reuter fixed point.
Yet, the ``weak-gravity'' region appears preferable both from a phenomenological point of view as well as regarding
the aspect of controlling the truncation.

In \cite{Eichhorn:2017eht}, a comprehensive analysis of the conditions under which a weak-gravity bound exists for
quartic matter couplings was put forward.
The corresponding beta functions are quadratic in the matter coupling.
Here, we will analyze the case of beta functions that are cubic in the coupling.
The beta function for $\sigma$ falls into this category.

In the previous sections, we have found it convenient to choose a parametrization of the action where $G$ appears
in those terms arising from the Einstein-term, in the minimally coupled interactions and in the nonminimal vertex.
An alternative parametrization, where $G$ does not appear in the nonminimal vertex, is related by a transformation
of the basis in theory space.
Specifically, the difference between the two parametrizations lies in the redefinition \eqref{eq:redef}.
To explore the weak-gravity bound, the 
modified
parametrization is more suitable.
This parametrization allows us to test the response of one sector, the nonminimal one, to the strengthening of
metric fluctuations.
In this parametrization, the strengthening can most conveniently be encoded in an increase of $G$.
The same cannot be done in our original parametrization, as an increasing of $G$ simultaneously ``dials'' the strength
of the nonminimal interaction term.
In this alternative parametrization, the flow equations
for $G_{\psi}$ and $\sigma_{\rm mod}$ in the $n\!\!=\!\!2$ scheme take the form given in
Eqs.~\eqref{eq: betan=2} and \eqref{eq: BSigMod}.
In this parametrization, $\beta_{\sigma_{\rm mod}}$ is, up to corrections from the loop contributions to the anomalous
dimensions, cubic in $\sigma_{\rm mod}$.
In the perturbative approximation, i.e. neglecting the anomalous dimensions coming from the
scale-derivative of the regulator it becomes exactly cubic.
This motivates a more general analysis of the existence of the shifted Gaussian fixed point in cubic beta functions.

Consider a generic beta function cubic in the coupling $\gamma$, i.e.,
\begin{equation} \label{eq: polynomlai}
 \beta_{\gamma} = a + b \, \gamma + c \, \gamma^{2} + d \, \gamma^{3} ,
\end{equation}
where $a,b,c$ and $d$ are coefficients
that are functions of other couplings of the system.
$b$ in general also contains a contribution from the canonical dimension of $\gamma$.
The shifted Gaussian fixed point (sGFP) is defined as the fixed point that is a continuous deformation of
the free fixed point for $a \neq 0$.
In the case of gravitational systems, it is the effective strength of metric fluctuations that leads to this deformation.
As the effective strength of metric fluctuations increases, the critical exponent associated to $\gamma$ changes.
As $\theta = 0$ is associated to a double zero of a beta function, a change of sign of the critical exponent
of the sGFP is necessarily tied to a fixed-point collision.
Such collisions can (but need not, see, e.g., \cite{Eichhorn:2013zza,Eichhorn:2015woa} for exceptions)
lead to a loss of real fixed points.
Up until a possible collision, a canonically irrelevant coupling is therefore irrelevant at the sGFP.
For $d < 0$, the beta function can either feature one or three zeros.
For the former case, the zero comes with negative slope, i.e., it corresponds to a fixed point at which the coupling
is relevant.
Therefore, in the case of $d < 0$, the sGFP only exists if the beta function has three real zeros.
This condition is equivalent to demanding that the local minimum of $\beta_{\gamma}$ is negative and the local
maximum positive.
Expressed in terms of  the coefficients, this leads to
\begin{equation}
 c^{2} \geq 3 \, b \, d ,
 \quad
 \beta_{\gamma}(\gamma_{\text{max}}) > 0 ,
 \quad \text{and} \quad
 \beta_{\gamma}(\gamma_{\text{min}}) < 0 .
\end{equation}
Given any cubic beta function, this set of conditions can be checked.
Regions in the gravitational parameter space where these conditions are violated do not allow for the sGFP to exist.
\begin{figure}[!t]
  \centering
  \includegraphics[width=\linewidth]{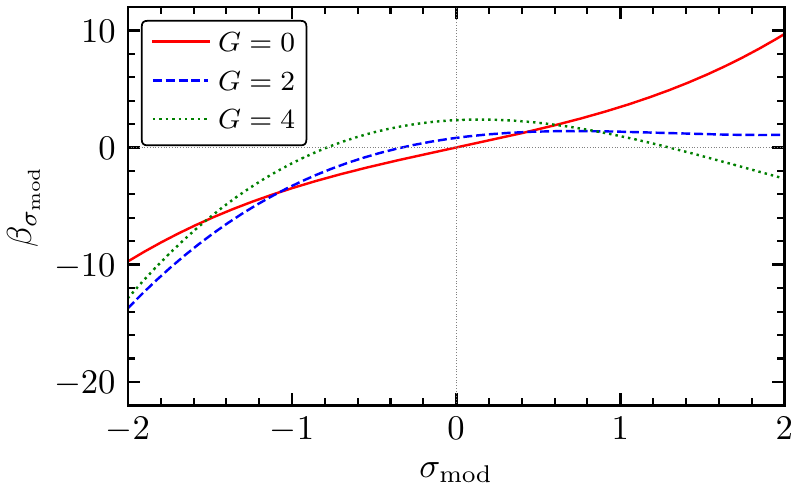}
  \caption{
  We show $\beta_{\sigma_{\rm mod}}$ after rescaling $\sigma_{\rm mod}$, such that $G$ does not appear in the nonminimal
  vertex, at $\lambda_{2} = 0.2$, $\lambda_{3} = 0$ with $G = 0$ (red, solid), $G = 2$ (blue, dashed) and
  $G = 4$ (green, dotted).
  } \label{fig: betasigmaresc}
\end{figure}

The case of $d > 0$ is more involved, since, depending on the specifics of the system, there can be either three,
or only one real zero point of $\beta_{\gamma}$.
In the latter case, the coupling $\gamma$ is irrelevant at two of the zeros.
Hence it is not clear from the values of the parameters, which of them is the sGFP.
To establish this requires tracking the sGFP all the way into the GFP as $a \rightarrow 0$.
Moreover, as the values of the parameters in the beta function are changed, the case with one zero can turn into the
case with three zeros.
Overall we conclude that for $d > 0$ there does not appear to be a simple criterion that allows to infer the existence
of the sGFP.
Instead, the sGFP has to be tracked explicitly starting from $a = 0$ to determine whether it exists in a given range of
parameter space.
Interestingly, in the asymptotic-safety literature, \cite{Gies:2016con} constitutes an example of the $d>0$ case.

We now turn our focus back to the coupling $\sigma_{\rm mod}$, with beta function
shown in Fig.~\ref{fig: betasigmaresc}.
In the case $G = 0$, the beta function features a Gaussian fixed point, as expected.
In accordance with symmetry arguments, this fixed point is shifted away from zero for $G \neq 0$, giving rise to an sGFP.
At the sGFP, $\sigma_{\rm mod}$ is shifted further into irrelevance.
This suggests indeed that canonical power counting is a suitable guiding principle to determine which operators
are relevant or irrelevant at the Reuter fixed point.

\begin{figure}[!t]
  \centering
  \includegraphics[width=\linewidth]{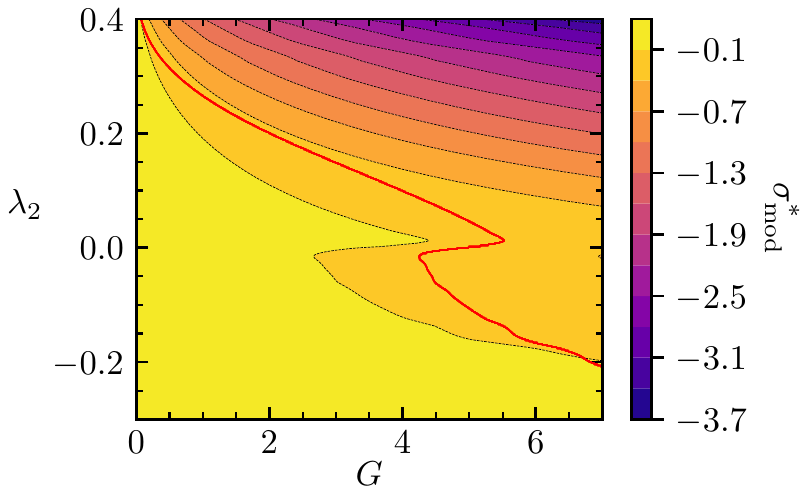}
  \caption{
  Value of the sGFP for the nonminimal coupling $\sigma_{\rm mod}$ at $\lambda_{3} = 0$, as a function of the gravitational
  parameters $G$ and $\lambda_{2}$.
  The existence of a real sGFP in the whole plot-range implies the absence of a weak-gravity bound in the same region.
  The red line indicates where one of the anomalous dimensions
  reaches
  $\eta = 2$.
  } \label{fig: MaxMinPosRescal}
\end{figure}

Following our analysis above, exploring whether the sGFP exists everywhere in the gravitational parameter space is
best done by explicitly computing the value of the sGFP as a function of the other couplings of the system.
Fig.~\ref{fig: MaxMinPosRescal} shows the value of the sGFP as a function of the gravitational parameters
$G$ and $\lambda_{2}$ at a fixed value of $\lambda_{3}$.
\begin{figure}[!t]
  \centering
  \includegraphics[width=\linewidth]{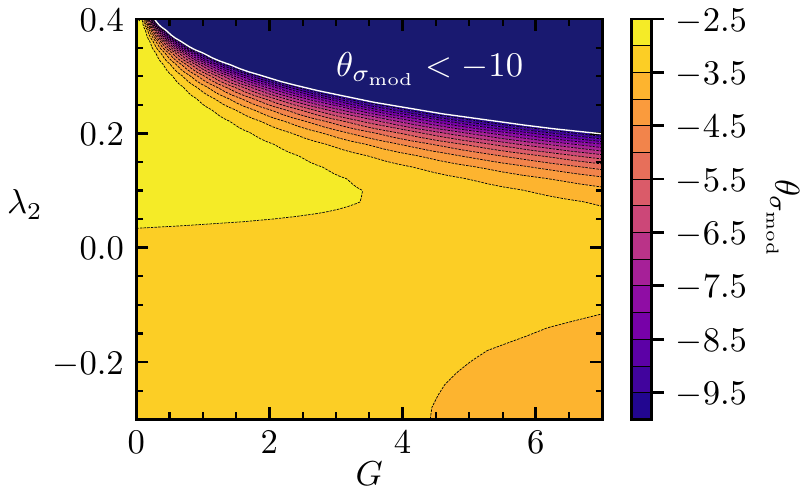}
  \caption{
  Values of the critical exponent $\theta_{\sigma_{\rm mod}}$ at the sGFP as a function of the gravitational parameters $G$
  and $\lambda_{2}$.
  } \label{fig: Thetas}
\end{figure}
There is no indication for a weak-gravity bound in the shown region.
Fig.~\ref{fig: MaxMinPosRescal} only covers a bounded region in theory space.
Beyond the red line in Fig.~\ref{fig: MaxMinPosRescal}, i.e., for larger values of $G$ and $\lambda_{2}$,
at least one of the anomalous dimensions violates $\eta \leq 2$, such that the truncation is not reliable any more.
Even though the anomalous dimensions must increase even further in order to flip the sign of diagrams contributing
to the flow, $\eta=2$ is the point where the regulator bound discussed in \cite{Meibohm:2015twa} is violated.
For more negative values of $\lambda_{2}$, the effective strength of gravity fluctuations decreases, cf.~the discussion
in Sec.~\ref{sec:Nfgtr1}.
As the weak-gravity bound is expected to be induced by strong gravity fluctuations, we do not expect the sGFP to vanish
into the complex plane for more negative values of $\lambda_{2}$.
Fig.~\ref{fig: Thetas} shows the values of the critical exponent $\theta_{\sigma_{\mathrm{mod}}}$ at the sGFP as a function
of the gravitational parameters.
We observe that the critical exponent at the sGFP is almost everywhere shifted further into
irrelevance starting from the canonical scaling $\theta_{\sigma_{\mathrm{mod}}} = -3$ at $G = \lambda_{2} = 0$.
This signals that the system is actually driven away from a fixed-point collision.
We conclude that in the region where our truncation is expected to provide robust results, there are no indications
for a weak-gravity bound.
Accordingly, the inclusion of the nonminimal interaction $\sigma_{\mathrm{mod}}$ does not lead to new constraints on
the microscopic gravitational parameter space. Thus, the asymptotic-safety scenario passes a nontrivial test:
The presence of a nonzero interaction, which has been neglected in previous studies, is innocuous in that its inclusion
does not impose new constraints and only results in subleading corrections to fixed-point values from previous studies.
\section{Asymptotic safety for more flavors}\label{sec:Nfgtr1}
\begin{figure*}[!t]
  \centering
  \includegraphics[width=\textwidth]{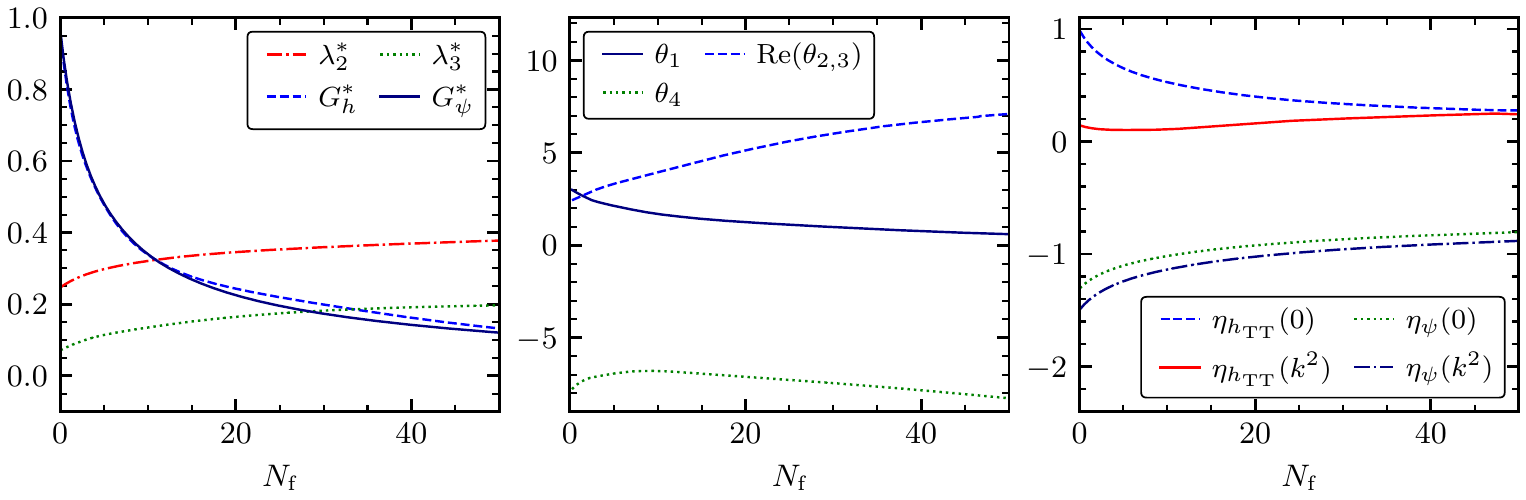}
  \includegraphics[width=\textwidth]{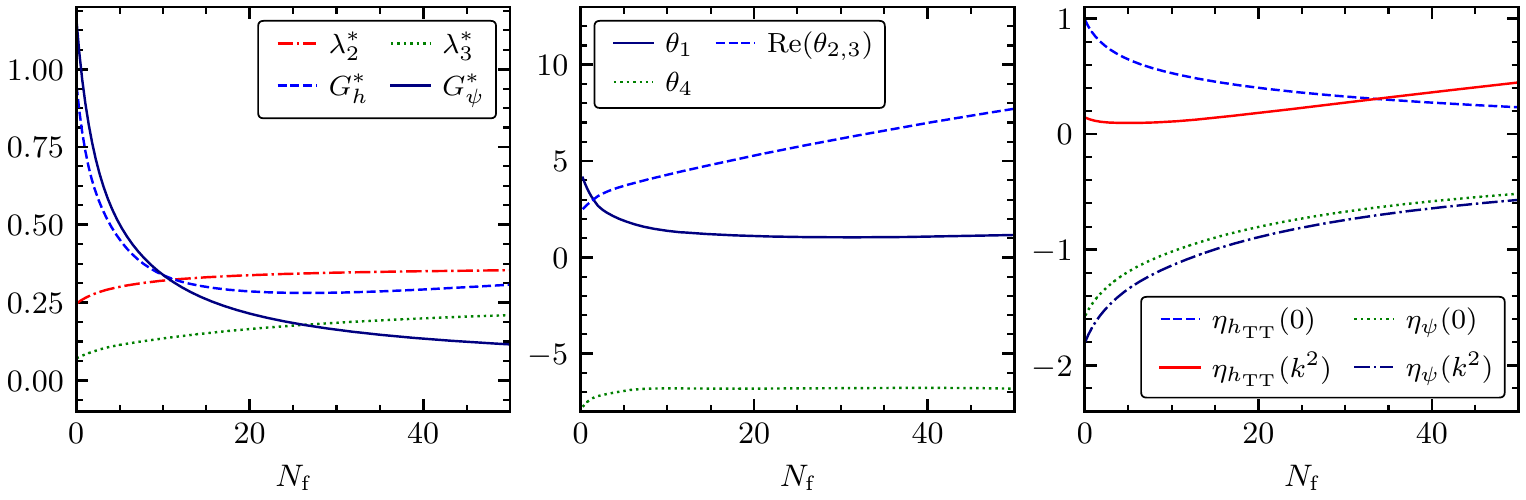}
  \caption{Upper panels: Fixed-point structure  as a function of $N_{\rm f}$ in $n\!\!=\!\!1$ projection.
  Lower panels:
  Fixed-point structure for the system in $n\!\!=\!\!2$ projection with $\sigma=0$ as function of the number of fermions
  $N_{\rm f}$.}\label{fig:TrilFP}
\end{figure*}
\begin{figure*}[!t]
  \centering
  \includegraphics[width=\textwidth]{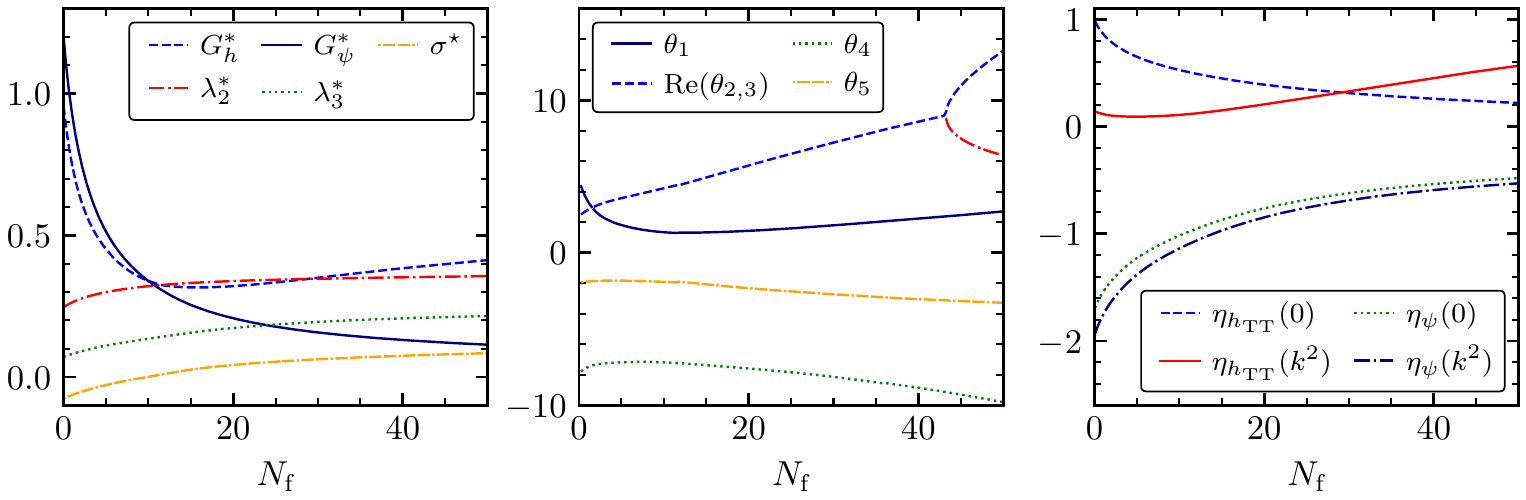}
  \caption{Fixed-point structure for the system with $n\!\!=\!\!2$ projection with $\sigma\neq 0$
  evaluated $\Gamma^{(h\bar{\psi}\psi)}$ as function of the number of fermions $N_{\text{f}}$.
  }\label{fig:TrilSFP}
\end{figure*}
\begin{figure*}[!t]
  \centering
  \includegraphics[width=\textwidth]{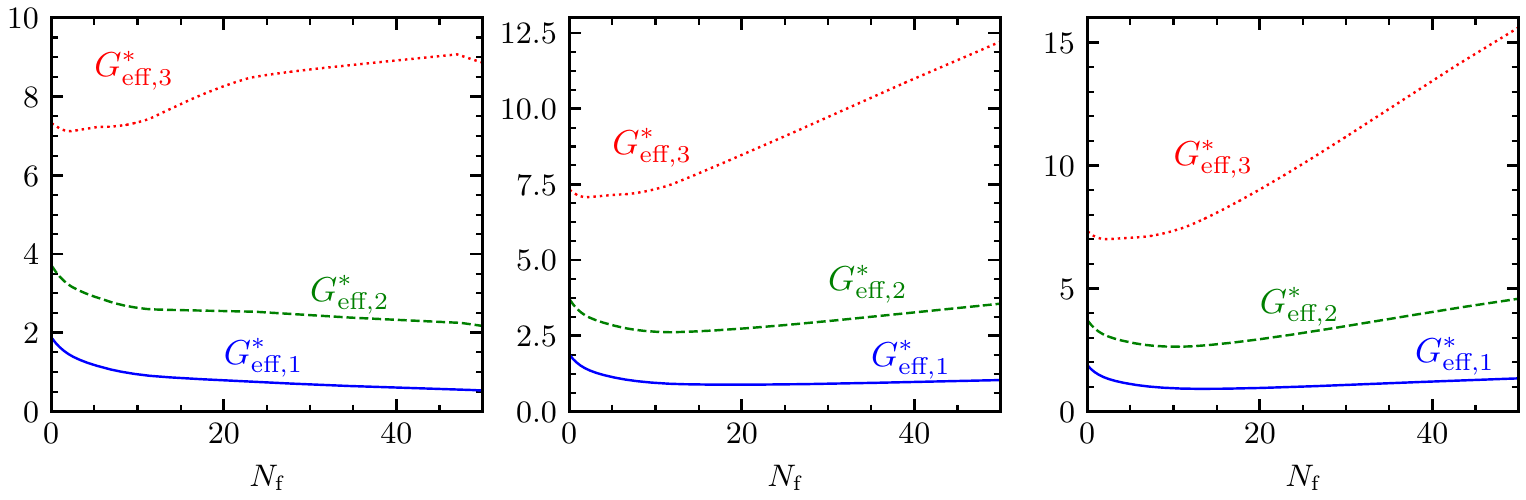}
  \caption{Value of the effective gravitational coupling $G_{\text{eff},\,n}^*$ for different values of $n$.
  Left panel: $n\!\!=\!\!1$ projection,
  central panel: $n\!\!=\!\!2$ with $\sigma=0$ and
  right panel: $n\!\!=\!\!2$ with $\sigma\neq0$.
  }\label{fig: Geff}
\end{figure*}

A key question on the asymptotic-safety scenario is whether it is compatible with arbitrary matter models or whether
it places restrictions on the matter sector.
If this was not the case, there would be a huge ``landscape" for asymptotic safety.
This would clearly make it much more challenging to confront the asymptotic-safety scenario with data,
as experimental data are only available at energies where the fixed-point scaling itself is not yet detectable.
If asymptotically safe gravity is only compatible with a very small set of matter models,
then one can hope for it to be ruled out by particle-physics data.
To establish which models lie in the asymptotically safe regime and which lie in the ``swampland",
the interaction structure of models also has to be taken into account.
Here, we restrict ourselves to the first step in the landscape/swampland classification, and ask whether
a bound exists on the number of fermion flavors compatible with a fixed point in our truncation.
Note that the realization of scale invariance in the UV relies on a delicate balance of competing effects of
quantum fluctuations of different fields.
Accordingly one might expect the asymptotically safe region to be a tiny ``island".

Further, the gravitational fixed-point values are a key input in determining whether asymptotically safe gravity
could provide accurate ``retrodictions'' of Standard Model couplings
\cite{Eichhorn:2017ylw,Eichhorn:2017lry,Eichhorn:2018whv}.
These fixed-point values depend on the number of matter fields, and accordingly an accurate determination
at $N_{\rm f}=22.5$ (SM) or $N_{\rm f}=24$ (SM+$\nu$'s) is required.
Our study is also a step towards reducing the systematic error of the gravitational fixed-point values.

There are compelling indications for a screening effect of fermions on the running Newton coupling,
as it is seen in perturbative studies \cite{Kabat:1995eq,Larsen:1995ax},
background studies \cite{Dona:2012am,Dona:2013qba,Eichhorn:2016vvy} and fluctuation field studies \cite{Meibohm:2015twa},
including the present one for $N_{\mathrm{f}} > 1.5$, cf.~Eq.~\eqref{eq: GpsiDeriv} and Eq.~\eqref{eq: GhDeriv}
in App.~\ref{sec:appderbeta} and Tab.~\ref{tab: BigTable}, where $A_{{\rm f}, G_{\psi}}>0$ for $N_{\rm f}>1.5$.

For the fluctuation system, the numerical prefactor of the $G^2\, N_{\mathrm{f}}$-term in $\beta_{G_{\psi}}$
and $\beta_{G_h}$ is significantly smaller than for the background system,
where it is $A_{\mathrm{f},G_B}/A_{G_B} \approx 0.04$ for Dirac fermions \cite{Dona:2013qba},
with the $A$'s defined in analogy to Eq.~\eqref{eq:betaA}.
In contrast, $A_{\mathrm{f},G_{\psi}}/A_{G_{\psi}} \approx 8.5 \cdot 10^{-4}$.
Accordingly, the Reuter fixed point in the fluctuation system changes much slower as $N_{\mathrm{f}}$ is increased.
Up to values of $N_{\mathrm{f}}$ in the range $N_{\mathrm{f}}\approx 20...30$, only slight changes are observed
in the system.
To understand the observed slight changes, the effect of fermions on $\lambda_2$ and $\lambda_3$ needs to be taken
into account.
This is the point where background calculations and fluctuation calculations differ:
As a function of increasing $N_{\mathrm{f}}$, the fixed-point value for the background cosmological constant becomes
increasingly negative, see, e.g., \cite{Dona:2013qba}.
At the same time, the fluctuation quantities $\lambda_2^{\ast}, \lambda_3^{\ast}$ both become more positive.

Our results are shown in Fig.~\ref{fig:TrilFP} for $\sigma=0$ and Fig.~\ref{fig:TrilSFP}.
We observe a growth of $\lambda_2^{\ast}$ and $\lambda_3^{\ast}$.
As $\lambda_2$ grows, it enhances the effective strength of metric fluctuations in the system of beta functions at hand.
The contribution of gravitons comes with $1/(1-2\lambda)^{\#}$, where $\#=1,2$ in our case.
Therefore a slight growth in $\lambda_2^{\ast}$ strengthens the metric contribution to $\beta_{G_{\psi/h}}$ sufficiently
to \emph{overcompensate} the effect of fermions, such that the fixed-point values in $G_{\psi/h}$ decrease as a function
of $N_{\mathrm{f}}$ up to $N_{\mathrm{f}} \approx 20-30$.
We caution that such threshold effects are regulator dependent, and accordingly it is an interesting open question
how a similar self-stabilization of the gravitational system could be encoded in beta functions with a different
choice of regulator.
For a related discussion, see \cite{Christiansen:2017cxa}.

At $N_{\mathrm{f}} \approx 20-30$, we observe hints for the onset of a more strongly coupled regime in our truncation.
For instance, a subset of the critical exponents deviates further from the canonical values.
This goes hand in hand with a more significant impact of $\sigma$, i.e., the fixed-point results with $\sigma$
start to deviate more significantly from those without.
Within the minimally coupled system, the two projections, shown in Fig.~\ref{fig:TrilFP} begin to show slight differences.
In particular, a difference between $G_h^{\ast}$ and $G_{\psi}^{\ast}$ begins to develop in the $n=2$ projection.
Moreover, effective universality for $G_h^{\ast}$ and $G_{\psi}^{\ast}$ appears to be lost in this regime.
That this corresponds to a more strongly-coupled regime can also be inferred by exploring the
effective strength of metric fluctuations. It is for instance encoded in the quantity
\begin{equation}
 G_{\mathrm{eff},\,n} = \frac{G}{(1-2\lambda_2)^n}.
\end{equation}
In the present projection, $\beta_{G_{\psi}}$ and $\beta_{G_h}$  are sensitive to $G_{\rm{eff}, \, 1/2}$.
Higher powers of $(1-2\lambda_2)$ can play a role in some diagrams.
As $\lambda_2$ grows towards the pole at $\lambda_2=1/2$, the effective strength of metric fluctuations can grow even
if $G$ itself does not grow.
We observe that the $G_{\rm eff}$ decrease as a function of $N_{\rm f}$ until about $N_{\rm f}\approx 20$,
when an increase is observed.
For the truncation without $\sigma$ in $n\!\!=\!\!1$ projection the increase only occurs for the higher-order $G_{\rm eff}$,
cf.~Fig.~\ref{fig: Geff}.

\begin{figure}[t!]
  \includegraphics[width=\linewidth]{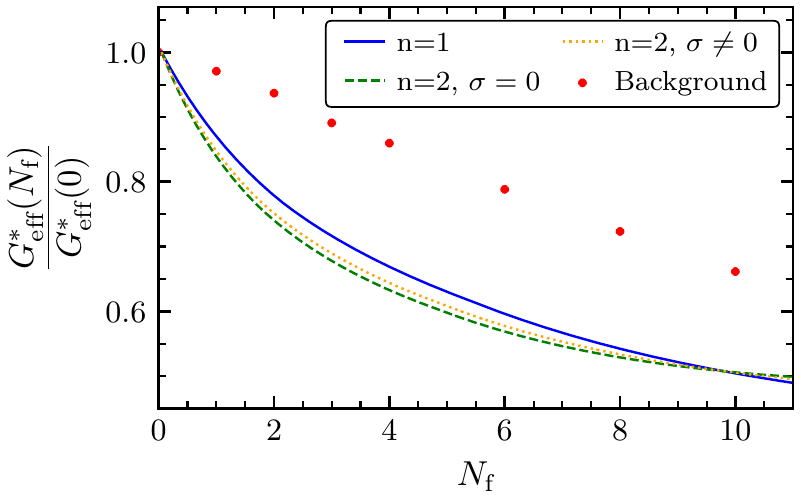}
  \caption{We show the value of $G_{\mathrm{eff},\,1}^*$ for different truncations as a function of the fermion number.
  The values for the background system are taken from \cite{Dona:2013qba}.
  }\label{fig:MMGeff}
\end{figure}

Taken together, this indicates a need to extend the truncation for a quantitatively reliable determination of fixed-point
properties in the regime beyond $N_{\rm f}\approx 20-30$.
A continuation of the Reuter fixed point also exists in this regime in our study.
Yet, a conservative estimation of the reliability of truncations suggests that extensions are necessary to
reliably probe its existence and possible properties.
Nevertheless, we highlight that our results could be interpreted to hint at the existence of a fixed point also at large
$N_{\rm f}$, which could be relevant for asymptotic safety in matter models \cite{Litim:2014uca,Molinaro:2018kjz}.
Accordingly, a scenario in which a crossover from asymptotically safe fixed-point scaling with gravity to asymptotic safety
without gravity determines high-energy physics, might potentially be realizable.
Such a setting requires further investigation.
Finally, we point out that $G_{\rm eff,1}$ exhibits similar behavior in the background- as in the fluctuation system
for $N_{\rm f}=1...10$.
In this region, $G_{\rm eff,1}$ falls for both systems, cf.~Fig.~\ref{fig:MMGeff}, although the underlying mechanisms at
the level of  fixed-point values differ.
This could be interpreted as a hint that at the level of physically relevant combinations of couplings,
the background- and fluctuation system could behave similarly.
\section{Conclusions and outlook}\label{sec:conclusions} 
\subsection{Key finding: A robust fermion-gravity fixed point for finitely many fermions}
In this paper, we have zoomed in on the microscopic dynamics of gravity and fermions.
Going beyond previous studies in the literature, we have explored a new direction in the space of couplings:
Nonminimal derivative couplings for fermions are expected to be present at the Reuter fixed point according
to symmetry arguments \cite{Eichhorn:2017eht}.
Here, we have included the leading-order one of this family of couplings, and confirmed that it cannot be consistently
set to zero.
Accordingly its inclusion constitutes a nontrivial test of asymptotically safe gravity.
In particular, we find good indications for a robust continuation of the Reuter fixed point from zero to finite
fermion number $N_{\rm f}\approx 20-30$.
Specifically, the inclusion of the canonically irrelevant, nonminimal coupling $\sigma$ has very little impact
on properties of the system determined in smaller truncations.
Most importantly, the introduction of $\sigma$ adds another irrelevant direction at the interacting fixed point.
Hence, our study provides further evidence for the small finite dimensionality of the UV-critical hypersurface of
the Reuter fixed point also in the presence of matter.
Our results further reinforce the scenario that asymptotically safe gravity could be near-perturbative,
implying that the spectrum of higher-order critical exponents follows a near-canonical scaling.
Moreover, we find further hints for a possible near-perturbative nature of asymptotic safety by comparing two
``avatars'' of the Newton coupling.
These are expected to exhibit a near-equality of fixed-point values, referred to as effective universality
\cite{Eichhorn:2018akn,Eichhorn:2018ydy}, in a near-perturbative regime.
Our investigation also includes a detailed analysis of the momentum-dependence of the flow, highlighting
the small impact of the nonminimal coupling in terms of a suppressed higher-momentum dependence of the
graviton-fermion vertex.

Beyond $N_{\rm f}\approx 20-30$, we find indications that further extensions of the truncation are required.
In this regime, the inclusion of $\sigma$ as well as changes in the projection scheme have an appreciable impact on
the fixed-point properties.
Moreover, critical exponents deviate further from canonical scaling.
To reliably test whether the Reuter fixed point can be extended to much higher fermion numbers,
extensions of the truncation are required to reach quantitatively robust control of the system at
$N_{\rm f} \gtrapprox 20-30$.
\begin{figure}[t!]
 \includegraphics[width=\linewidth]{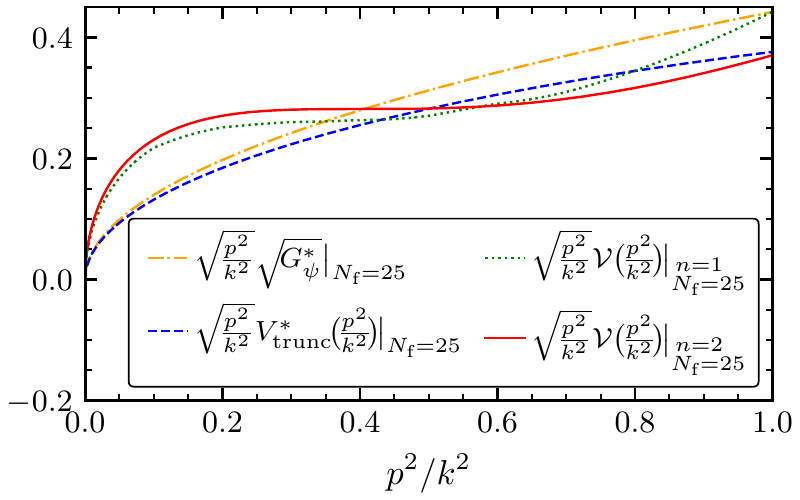}
 \caption{
 Momentum dependence of the graviton-fermion vertex for $N_{\mathrm{f}}=25$ fermions for the $n\!\!=\!\!1$
 and the $n\!\!=\!\!2$ projection including $\sigma$.
 }
 \label{fig: MomDep25Ferms}
\end{figure}
\subsection{Outlook}
For values of $N_{\mathrm{f}}>30$, the large deviation of the critical exponents from canonical scaling indicates
the need for further extensions of the truncation.
In principle, many different directions in theory space are available for such an extension.
Accordingly it is highly desirable to obtain an educated guess, which direction is the most likely to provide
an important step towards apparent convergence of the results.
We have already observed that the system is rather stable under the inclusion of the coupling $\sigma$.
Specifically, it is useful to consider the momentum-dependence of the flow at $N_{\mathrm{f}}>1$.
Performing a similar comparison as for the case $N_{\mathrm{f}}=1$ in Fig.~\ref{fig: MomDep25Ferms}, we find that the
$n\!\!=\!\!2$ approximation with $\sigma$ captures the momentum-dependence of the flow more accurately
than the approximation without $\sigma$ ($n\!\!=\!\!1$),  cf.~Fig.~\ref{fig: MomDep25Ferms}.
Therefore, our extension of the truncation is an important step towards quantitative precision for $N_{\rm f}\approx 25$.
A further extension to $p^5$ terms in the momentum-dependence also appears indicated by the results,
cf.~Fig.~\ref{fig: MomDep25Ferms}.

Further contributions from the matter sector to $\beta_{G_{\psi}}$ might be relevant at large $N_{\mathrm{f}}$.
In particular, this could include induced, chirally symmetric four-fermions interactions \cite{Eichhorn:2011pc}.
In \cite{Eichhorn:2017eht} a suppression of the contribution of these induced interactions to other beta functions
was observed at $N_{\mathrm{f}}=1$.
Yet, an explicit study of the impact of these interactions to $\beta_{G_{\psi}}$ is still outstanding.
Moreover, a scaling of this contribution with $N_{\mathrm{f}}$ could be possible,
and could therefore enhance the corresponding effects at large $N_{\mathrm{f}}$.
\begin{figure}[t!]
  \includegraphics[width=\linewidth]{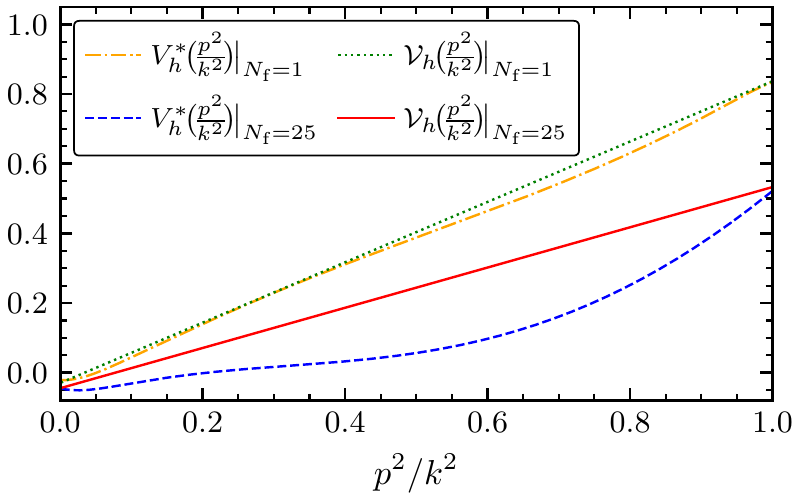}
  \caption{Momentum dependence for the graviton three-point function evaluated at the corresponding fixed point.}
  \label{fig: MomDepOutlook}
\end{figure}

Fig.~\ref{fig: MomDepOutlook} shows the momentum dependence of the graviton three-point vertex at
$N_{\mathrm{f}}=1$ and $N_{\mathrm{f}}=25$.
For the former case, it is well approximated by a polynomial up to first order in $p^2$.
This is not the case for $N_{\mathrm{f}}=25$.
There, the momentum dependence shows a clear $p^4$ dependence which is not captured by the $n=2$
projection at $p^2=0$ and $p^2=k^2$ which is the approximation we work in for $G_h$ in this paper.
Accordingly, at large fermion numbers, the $n\!\!=\!\!2$ projection introduces an error.
This might be the reason for the deviation of the fixed-point values of both avatars of the Newton coupling,
visible, e.g., in Fig.~\ref{fig:TrilSFP}.
Therefore, an extension of the truncation in the pure-gravity direction is indicated, in order to obtain more reliable
results for the behavior at $N_{\mathrm{f}}>30$.
In Fig.~\ref{fig: MomDepOutlook} we observe that a polynomial up to second order in $p^2$ might capture
the full momentum dependence of the graviton vertex.
Therefore the inclusion of $R^2$ and $R_{\mu\nu}^2$ would be the first step towards a more reliable
large-$N_{\mathrm{f}}$ behavior.
One might interpret our observation as the generation of a sizable $R_{\mu\nu}R^{\mu\nu}$ and/or $R^2$
term by fermionic fluctuations.
This is particularly intriguing, as the corresponding change in the effective graviton propagator might play a role in
generating quantum-gravity contributions to Standard Model beta functions of the appropriate size to be phenomenologically
viable \cite{Eichhorn:2017eht}.

In summary, we aim at reaching quantitative convergence of fixed-point values at finite fermion numbers.
A central motivation is that this is a key input for investigations into the phenomenology of asymptotic safety,
both in particle physics, see, e.g., \cite{Eichhorn:2018yfc} as well as in cosmology, see, e.g., \cite{Bonanno:2017pkg}.

\emph{Acknowledgements:} We thank J.~M.~Pawlowski and M.~Reichert for discussions.
A.~E.~and M.~S.~are supported by the DFG under grant no.~Ei/1037-1.
A.~E.~is grateful to the African Institute for Mathematical Sciences (AIMS) Ghana for hospitality during the final stages
of this work.
S.L.~is supported by the DFG Collaborative Research Centre "SFB 1225 (ISO-QUANT)".
\appendix
\section{Dimensionless form of the Wetterich equation}\label{sec:app:flow}
In this section we explain in detail, how one can cast the flow equation into a completely dimensionless
and $k$-independent form.
This leads us to a compact expression containing every beta function of the system and provides a formally exact
equation underlying all fixed-point searches in truncations.
After that, we give an explicit example on how to extract specific beta functions.
\subsection{Generic case}
In order to get the general picture we will, for now, mostly use the DeWitt notation with super indices $A$,
the super field $\Phi^{A}$,
\begin{align}
 (\Phi^{A}) = \big( h_{\mu \nu}(x) , \psi^{i}(x), \bar{\psi}^{i}(x), c^{\mu}(x), \bar{c}_{\mu}(x) \big),
\end{align}
and the effective action $\Gamma_{k}$ without reference to any truncation.
For example, in this language the vertex expansion of the effective action reads
\begin{align}
 \Gamma_{k}[\Phi;\bar{g}]
 ={}& \sum\limits_{n = 0}^{\infty} \frac{1}{n!} \Gamma_{k \, A_{1} \ldots A_{n}}^{(n)}[0;\bar{g}]
 \Phi^{A_{n}} \ldots \Phi^{A_{1}},
\end{align}
where the $\Gamma_{k}^{(n)}$ are the functional derivatives of the effective action with respect to the fields,
\begin{align}
 \Gamma_{k \, A_{1} \ldots A_{n}}^{(n)}[\Phi;\bar{g}]
 = \Gamma_{k} [\Phi;\bar{g}]
 \frac{\overleftarrow{\delta}}{\delta \Phi^{A_{1}}} \ldots \frac{\overleftarrow{\delta}}{\delta \Phi^{A_{n}}}.
\end{align}
Note the order of the indices and fields, which is important to keep in mind for the Grassmann-valued quantities.
When performing RG steps, i.e., when changing $k$, all $n$-point functions change.
However, some parts of that we can absorb by a simple rescaling of the field.
To capture this, let us define the (Grassmann even) generalized wave-function renormalization $\mathcal{Z}_{k}$
such that the second functional derivative of $\Gamma_{k}$ agrees with the standard kinetic term,
\begin{align}
 \frac{1}{2} {\rm Tr} \big( \Gamma_{k}^{(2)}[0;\bar{g}] \Phi \Phi^{\rm T} \big)
 ={}& \frac{1}{2} {\rm Tr} \big( \mathcal{Z}_{k}[\bar{g}]^{\rm T} {\rm Kin}_{k}[\bar{g}] \mathcal{Z}_{k}[\bar{g}]
 \Phi \Phi^{\rm T} \big),
\end{align}
where ${\rm Kin}_{k}$ is the standard kinetic operator.
The wave-function renormalization is a matrix in field space.
Thus, for practical purposes we assign a factor of $\mathcal{Z}_{k}$
instead of $\mathcal{Z}_{k}^{1/2}$ to each field.
Further, note the remaining $k$-dependence in ${\rm Kin}_{k}$, which is present due to the following reason:
In principle we can bring $\Gamma_{k}^{(2)}$ into any desired form, however as $\mathcal{Z}_{k}$
is supposed to be a rescaling it must not vanish.
Therefore we cannot remove, e.g., mass poles, whose location we still need to follow while changing $k$.

One important ingredient for a full RG step is the rescaling of the (background) spacetime.
This is usually implemented by a rescaling of the coordinates, $x^{\mu} \to k^{-1} x^{\mu}$.
Here we will follow a different route:
as the diffeomorphism invariant (background) distance is
$\mathrm{d} s^{2} = \bar{g}_{\mu \nu} \mathrm{d} x^{\mu} \mathrm{d} x^{\nu}$,
we can implemented the rescaling by reparametrizing the background metric $\bar{g}$ in terms of a dimensionless
metric $\hat{\bar{g}}$,%
\footnote{%
In this appendix we use a hat to indicate dimensionless quantities, e.g., $\hat{\bar{g}}$,
whereas dimensionful quantities ``wear'' no additional symbols,
in order to avoid confusion of dimensionful quantities with the bar of background quantities, e.g., $\bar{g}$.
}
\begin{align} \label{eq:bar_g_k}
 \bar{g}_{\mu \nu} \to \bar{g}_{k \, \mu \nu}[\hat{\bar{g}}] = k^{-2} \hat{\bar{g}}_{\mu \nu}.
\end{align}
In particular this implies that we treat coordinates as dimensionless, while the metric carries the dimensionality.
For example the operator $\bar{\slashed{\nabla}}$ still scales with $k$,
however, the $k$ now does not arise from the dimensionless $\bar{\nabla}_{\mu}$, but from the
dimensionful $\bar{\gamma}^{\mu} = \bar{g}^{\mu \nu} \bar{\gamma}_{\nu}$,
\begin{align}
 \{ \bar{\gamma}_{\mu} , \bar{\gamma}_{\nu} \} = 2 \bar{g}_{\mu \nu}
 \quad \to \quad
 \{ k^{-1} \hat{\bar{\gamma}}_{\mu} , k^{-1} \hat{\bar{\gamma}}_{\nu} \} = 2 k^{-2} \hat{\bar{g}}_{\mu \nu}.
\end{align}
The advantage of this assignment of dimensionality is twofold.
Firstly this fits better to the diffeomorphic nature of gravity, where one can choose coordinates as one pleases.
For instance, for spherical coordinates, it is obvious that we do not want the angles to carry any dimension.
Yet, if we insist on the radial coordinate to carry a dimension, then
the metric tensor would not have a homogenous dimension, as the purely radial component would be dimensionless,
while the angular components of the metric do carry nonzero dimensionality.
The second advantage is that we can make the rescaling of the spacetime manifest in the effective action, by using
$k^{-2} \hat{\bar{g}}$ instead of $\bar{g}$ as the (background) metric argument of $\Gamma_{k}$.
This would be somewhat more complicated if we insisted on rescaling the coordinates.

Considering the rescaling of the (background) spacetime, we define the (in field space) diagonal operator $\mathcal{K}_{k}$,
such that it accounts for the canonical mass dimension of the fields.
We can read off the canonical scaling with $k$ from the behavior of the kinetic term under a rescaling
of the (background) metric,
\begin{align}
 {\rm Kin}_{k}[k^{-2} \hat{\bar{g}}]
 ={}& \mathcal{K}^{-1}_{k} \widehat{{\rm Kin}}_{k}[\hat{\bar{g}}] \mathcal{K}^{-1}_{k},
 \quad \mathcal{K}_{k}^{\rm T} = \mathcal{K}_{k},
\end{align}
where $\widehat{\rm Kin}_{k}$ is the dimensionless kinetic operator with dimensionless couplings (mass poles).
The scale derivative of $\mathcal{K}_{k}$ then gives a factor $\mathcal{N}$ corresponding to the canonical mass dimension,
\begin{align}
 \partial_{t} \mathcal{K}_{k} = \mathcal{N} \mathcal{K}_{k} = \mathcal{K}_{k} \mathcal{N}.
\end{align}
With the above definitions we can parametrize the dimensionful field $\Phi$ in terms of the dimensionless field $\hat{\Phi}$,
\begin{align} \label{eq:Phi_k}
 \Phi \to \Phi_{k}[\hat{\Phi};\hat{\bar{g}}] = \mathcal{Z}_{k}^{-1}[k^{-2} \hat{\bar{g}}] \mathcal{K}_{k} \hat{\Phi}.
\end{align}
In our case this translates into the following canonical powers of $k$
\begin{align}
 {}& (\tensor{{\mathcal{K}_{k}}}{^{A}_{B}} \hat{\Phi}^{B})
 \\ \notag
 {}& \, = k^{\frac{d-6}{2}}
 \big( \hat{h}_{\mu \nu}(x), k^{\frac{5}{2}} \hat{\psi}(x), k^{\frac{5}{2}} \hat{\bar{\psi}}(x),
 k^{2} \hat{c}^{\mu}(x), k^{2} \hat{\bar{c}}_{\mu}(x) \big),
\end{align}
and corresponding canonical mass dimension $\mathcal{N}$
\begin{align}
 {}& (\tensor{\mathcal{N}}{^{A}_{B}} \hat{\Phi}^{B})
 \\ \notag
 {}& \, = \big( \tfrac{d-6}{2} \hat{h}_{\mu \nu}(x), \tfrac{d-1}{2} \hat{\psi}(x), \tfrac{d-1}{2} \hat{\bar{\psi}}(x),
 \tfrac{d-2}{2} \hat{c}^{\mu}(x), \tfrac{d-2}{2} \hat{\bar{c}}_{\mu}(x) \big).
\end{align}
Note the canonical scaling of $h_{\mu \nu} \sim k^{\frac{d-6}{2}} \hat{h}_{\mu \nu}$.
At first sight this seems different from the standard one of a bosonic field $\sim k^{\frac{d-2}{2}}$.
This ``different'' factor arises from the positioning of the indices in the kinetic term for $h_{\mu \nu}$,
\begin{align}
 \frac{1}{2} {\rm Tr} \big( {\rm Kin}_{k}[k^{-2} \hat{\bar{g}}] h h^{\rm T} \big)
 \sim k^{6-d} \int \!\! \sqrt{\hat{\bar{g}}} \, h_{\mu \nu} \, \hat{\bar{\Delta}} \, h_{\rho \sigma}
 \hat{\bar{g}}^{\mu \rho} \hat{\bar{g}}^{\nu \sigma}.
\end{align}
If we were to use $\tilde{h}^{\mu}_{\nu} = \bar{g}^{\mu \rho} h_{\rho \nu}$ instead of $h_{\mu \nu}$,
\begin{align}
 \frac{1}{2} {\rm Tr} \big( {\rm Kin}_{k}[k^{-2} \hat{\bar{g}}] \tilde{h} \tilde{h}^{\rm T} \big)
 \sim k^{2-d} \int \!\! \sqrt{\hat{\bar{g}}} \, \tilde{h}^{\mu}_{\nu} \, \hat{\bar{\Delta}} \, \tilde{h}^{\nu}_{\mu},
\end{align}
we would find the usual $\sim k^{\frac{d-2}{2}}$.
One can easily see that this choice has no impact on any of the beta functions,
as every such factor can be reabsorbed into the wave-function renormalization.

Finally let us define the generalized anomalous dimension $\eta_{k}$, which captures the anomalous scaling of the fields,
i.e., the scaling not coming from $\mathcal{K}_{k}$ but from $\mathcal{Z}_{k}$,
\begin{align}
 \eta_{k}[\hat{\bar{g}}] = - 2 \mathcal{K}_{k}^{-1} \dot{\mathcal{Z}}_{k}[k^{-2} \hat{\bar{g}}]
 \mathcal{Z}_{k}^{-1}[k^{-2} \hat{\bar{g}}] \mathcal{K}_{k}.
\end{align}
The full scaling of the dimensionful field $\Phi_{k}$ then reads
\begin{align}
 \mathcal{K}_{k}^{-1} \mathcal{Z}_{k}[k^{-2} \hat{\bar{g}}] \partial_{t} \Phi_{k}[\hat{\Phi}; \hat{\bar{g}}]
 = (\tfrac{1}{2} \eta_{k}[\hat{\bar{g}}] + \mathcal{N}) \hat{\Phi}.
\end{align}
Using the above we can define the dimensionless effective action $\hat{\Gamma}_{k}$,
\begin{align}
 \hat{\Gamma}_{k}[\hat{\Phi}; \hat{\bar{g}}]
 = \Gamma_{k}[\mathcal{Z}_{k}^{-1}[k^{-2} \hat{\bar{g}}] \mathcal{K}_{k} \hat{\Phi}; k^{-2} \hat{\bar{g}}].
 \label{eq: dimlessgammak}
\end{align}
Next we choose a regulator $R_{k}$ of the form
\begin{align}
 R_{k}[\bar{g}]
 ={}& \mathcal{Z}_{k}^{\rm T}[\bar{g}] \mathcal{K}_{k}^{-1} \hat{R}[k^{2} \bar{g}]
 \mathcal{K}_{k}^{-1} \mathcal{Z}_{k}[\bar{g}],
 \label{eq: regul}
\end{align}
where $\hat{R}[\hat{\bar{g}}]$ is a dimensionless and $k$ independent regulator shape function.
This is in full agreement of the standard way of writing the shape-function as a function of a dimensionful Laplacian
divided by $k^2$.
Expressed in terms of the dimensionless Laplacian, the shape function has no $k$-dependence.

Now we can move on to the calculation of the flow of the dimensionless effective action,
which explicitly contains all beta functions of the considered system,
\begin{align}
 \beta_{\hat{\Gamma}_{k}}\big[\hat{\Gamma}_{k}; \hat{R}, \hat{\bar{g}} \big][\hat{\Phi}]
 = \partial_{t} \hat{\Gamma}_{k}[\hat{\Phi}; \hat{\bar{g}} ].
\end{align}
We start with the standard flow equation,
\begin{align}
 \dot{\Gamma}_{k}[\Phi; \bar{g}]
 = \frac{1}{2} {\rm STr} \, \big[ (\Gamma_{k}^{(2)}[\Phi; \bar{g}] + R_{k}[\bar{g}])^{-1}
 \dot{R}_{k}[\bar{g}] \big].
 \label{eq: floweq}
\end{align}
On the right hand side we can insert the functional derivative of the dimensionless effective action
(\ref{eq: dimlessgammak}) and our chosen regulator (\ref{eq: regul}), leading to
\begin{align} \label{eq: PropDimL}
 \big({}& \Gamma_{k}^{(2)}\big[ \Phi_{k}[\hat{\Phi};\hat{\bar{g}}]; \bar{g}_{k}[\hat{\bar{g}}] \big]
 + R_{k}\big[ \bar{g}_{k}[\hat{\bar{g}}] \big] \big)^{-1}
 \\ \notag
 {}& = \mathcal{Z}_{k}^{-1}[k^{-2} \hat{\bar{g}}] \mathcal{K}_{k}
 (\hat{\Gamma}_{k}^{(2)}[\hat{\Phi}; \hat{\bar{g}}] + \hat{R}_{k}[\hat{\bar{g}}])^{-1}
 \mathcal{K}_{k} \mathcal{Z}_{k}^{-1 \rm T}[k^{-2} \hat{\bar{g}}],
\end{align}
where $\bar{g}_{k}$ and $\Phi_{k}$ are given in Eqs.~\eqref{eq:bar_g_k} and \eqref{eq:Phi_k}.
In terms of dimensionless and $k$-independent quantities, the scale derivative of the regulator can be expressed via
\begin{align} \label{eq:rdot}
 \mathcal{K}_{k} {}& \mathcal{Z}_{k}^{-1 \rm T}[k^{-2} \hat{\bar{g}}] \dot{R}_{k}[k^{-2} \hat{\bar{g}}]
 \mathcal{Z}_{k}^{-1}[k^{-2} \hat{\bar{g}}] \mathcal{K}_{k}
 \\ \notag
 ={}& \!\!  - \! \big( \! \tfrac{1}{2}  \eta_{k}^{\rm T}[\hat{\bar{g}}] \! + \! \mathcal{N} \big)
 \hat{R}[\hat{\bar{g}}]
 \! - \! \hat{R}[\hat{\bar{g}}] \big( \! \tfrac{1}{2} \eta_{k}[\hat{\bar{g}}] \! + \! \mathcal{N} \big)
 \! - \! \partial_{q} \hat{R}[q^{\! -2} \hat{\bar{g}}]|_{q=1}.
\end{align}
In the last term we treated the $k$-derivative in terms of an auxiliary dimensionless parameter $q$.
This trick helps us making everything explicitly $k$ independent.
We use the $q$-derivative in order to capture the scaling with the metric.
In practice such a term essentially leads to derivatives with respect to the (dimensionless) momentum, e.g.,
\begin{align}
 \Delta[q^{-2} \hat{\bar{g}}] = q^{2} \Delta[\hat{\bar{g}}].
\end{align}
By comparison with the definition of the dimensionless effective action (\ref{eq: dimlessgammak}),
and applying the chain rule for derivatives, we observe
that the scale derivative of $\hat{\Gamma}_{k}$ decomposes into the diagrams,
coming from the standard flow equation, and the scaling of the fields:
\begin{align}
 \partial_{t} \hat{\Gamma}_{k} [\hat{\Phi}; \hat{\bar{g}}]
 \! = {}& \! \dot{\Gamma}_{k} \big[ \mathcal{Z}_{k}^{-1}[k^{-2} \hat{\bar{g}}] \mathcal{K}_{k} \hat{\Phi} ;
 k^{-2} \hat{\bar{g}}
 \big]
 \\ \notag
 {}& \!\!\! + \! \hat{\Gamma}_{k}^{(1)}[\hat{\Phi} ; \hat{\bar{g}}]
 ( \tfrac{1}{2} \eta_{k}[\hat{\bar{g}}] \! + \! \mathcal{N} ) \hat{\Phi}
 \! + \! \partial_{q} \hat{\Gamma}_{k}[\hat{\Phi} ; q^{\! -2} \hat{\bar{g}}]|_{q=1} .
\end{align}
Replacing the first line with the standard flow equation (\ref{eq: floweq}),
while using the expressions (\ref{eq: PropDimL}) and (\ref{eq:rdot}),
we are led to the flow equation for $\hat{\Gamma}_{k}$,
\begin{widetext}
\begin{align} \label{eq:full_beta_functional}
 \beta_{\hat{\Gamma}_{k}}[\hat{\Gamma}_{k}; \hat{R}, \hat{\bar{g}}][\hat{\Phi}]
 ={}& - \frac{1}{2} {\rm STr} \, \Big[ ( \hat{\Gamma}_{k}^{(2)}[\hat{\Phi} ; \hat{\bar{g}}]
 + \hat{R}[\hat{\bar{g}}] )^{-1} \Big( 
 \big( \tfrac{1}{2}  \eta_{k}^{\rm T}[\hat{\bar{g}}] + \mathcal{N} \big)
 \hat{R}[\hat{\bar{g}}]
 + \hat{R}[\hat{\bar{g}}] \big( \tfrac{1}{2} \eta_{k}[\hat{\bar{g}}] + \mathcal{N} \big)
 + \partial_{q} \hat{R}[q^{-2} \hat{\bar{g}}]|_{q=1}
 \Big) \Big]
 \\ \notag
 {}& + \hat{\Gamma}_{k}^{(1)}[\hat{\Phi} ; \hat{\bar{g}}] ( \tfrac{1}{2} \eta_{k}[\hat{\bar{g}}] + \mathcal{N} ) \hat{\Phi}
 + \partial_{q} \hat{\Gamma}_{k}[\hat{\Phi} ; q^{-2} \hat{\bar{g}}]|_{q=1}.
\end{align}
\end{widetext}
Note that we now have a completely $k$ independent flow equation, with a compact form for the fixed-point equation,
\begin{align}
 \beta_{\hat{\Gamma}_{k}}\big[ \hat{\Gamma}_{k}^{\ast}[\hat{R}, \hat{\bar{g}}]
 ; \hat{R}, \hat{\bar{g}} \big] = 0,
\end{align}
where $\hat{\Gamma}_{k}^{\ast}[\hat{R},\hat{\bar{g}}]$ is the fixed point action
as a function of the chosen regulator shape and the chosen (background) spacetime.

Within the vertex expansion scheme a point in theory space can be specified by choosing the
$\hat{\Gamma}_{k}^{(n)}[0;\hat{\bar{g}}]$ for $n \neq 2$
and choosing the remaining parameters (mass poles) in $\widehat{\rm Kin}_{k}[\hat{\bar{g}}]$, while
$\hat{\Gamma}_{k}^{(2)}[0; \hat{\bar{g}}]$ is then given by
\begin{align}
 \hat{\Gamma}_{k}^{(2)}[0;\hat{\bar{g}}]
 = \widehat{\rm Kin}_{k}[\hat{\bar{g}}].
\end{align}
The dimensionless effective action $\hat{\Gamma}_{k}$ then reads,
\begin{align} \notag
 \hat{\Gamma}_{k}[\hat{\Phi};\hat{\bar{g}}]
 ={}& \hat{\Gamma}_{k}[0;\hat{\bar{g}}]
 + \hat{\Gamma}_{k}^{(1)}[0;\hat{\bar{g}}] \hat{\Phi}
 + \tfrac{1}{2} {\rm Tr} \big( \widehat{\rm Kin}_{k}[\hat{\bar{g}}] \hat{\Phi} \hat{\Phi}^{\rm T} \big)
 \\
 {}& + \sum\limits_{n = 3}^{\infty} \frac{1}{n!} \hat{\Gamma}_{k \, A_{1} \ldots A_{n}}^{(n)}[0;\hat{\bar{g}}]
 \hat{\Phi}^{A_{n}} \ldots \hat{\Phi}^{A_{1}}.
\end{align}
The flow $\beta_{\hat{\Gamma}^{(n)}_{k}}$ of the vertices $\hat{\Gamma}_{k}^{(n)}[0;\hat{\bar{g}}]$ ($n \neq 2$)
is calculated by taking $n$ functional derivatives of equation \eqref{eq:full_beta_functional} with respect to $\hat{\Phi}$
and evaluate the expression at $\hat{\Phi} = 0$,
\begin{widetext}
\begin{align} \label{eq:flow_Gamma_n}
 \beta_{\hat{\Gamma}^{(n)}_{k}}[\hat{\Gamma}_{k};\hat{R} , \hat{\bar{g}}]_{A_{1} \ldots A_{n}}
 = {\rm Flow}^{(n)}_{A_{1} \ldots A_{n}}[\hat{\bar{g}}]
 + \sum\limits_{l=1}^{n} \hat{\Gamma}_{k \, A_{1} \ldots A_{l-1} B A_{l+1} \ldots A_{n}}^{(n)}[0;\hat{\bar{g}}]
 \tensor{(\tfrac{1}{2} \eta_{k}[\hat{\bar{g}}] + \mathcal{N})}{^{B}_{A_{l}}}
 + \partial_{q} \hat{\Gamma}_{k \, A_{1} \ldots A_{n}}^{(n)}[0; q^{-2} \hat{\bar{g}}].
\end{align}
where ${\rm Flow}^{(n)}$ is a short hand for the corresponding diagrams,
\begin{align}
 {\rm Flow}_{A_{1} \ldots A_{n}}^{(n)}[\hat{\bar{g}}] \!
 ={}& \! \tfrac{1}{2} {\rm STr} \big[ ( \hat{\Gamma}_{k}^{(2)}[\hat{\Phi} ; \hat{\bar{g}}]
 \! + \! \hat{R}[\hat{\bar{g}}] )^{-1} \big( 
 \!\! - \! \big( \tfrac{1}{2}  \eta_{k}^{\rm T}[\hat{\bar{g}}] \! + \! \mathcal{N} \big)
 \hat{R}[\hat{\bar{g}}]
 \! - \! \hat{R}[\hat{\bar{g}}] \big( \tfrac{1}{2} \eta_{k}[\hat{\bar{g}}] \! + \! \mathcal{N} \big)
 \! - \! \partial_{q} \hat{R}[q^{-2} \hat{\bar{g}}]|_{q=1}
 \big) \! \big] \! \frac{\overleftarrow{\delta}}{\delta \hat{\Phi}^{A_{1}}} \! \ldots \!
 \frac{\overleftarrow{\delta}}{\delta \hat{\Phi}^{A_{n}}} \! \Bigg|_{\hat{\Phi}=0} \! .
\end{align}
\end{widetext}
Furthermore the generalized anomalous dimension needs to be determined self-consistently,
by taking two functional derivatives of equation \eqref{eq:full_beta_functional} and evaluating at $\hat{\Phi} = 0$,
\begin{align} \label{eq:flow_eta}
 \beta_{\widehat{\rm Kin}_{k}}[\hat{\Gamma}_{k} ; \hat{R} , \hat{\bar{g}}]
 \! ={}& \! {\rm Flow}^{(2)}[\hat{\bar{g}}]
 + (\tfrac{1}{2} \eta_{k}^{\rm T}[\hat{\bar{g}}] + \mathcal{N})
 \widehat{\rm Kin}_{k}[\hat{\bar{g}}]
 \\ \notag
 {}& \!\!\! + \widehat{\rm Kin}_{k}[\hat{\bar{g}}] (\tfrac{1}{2} \eta_{k}[\hat{\bar{g}}] + \mathcal{N})
 + \partial_{q} \widehat{\rm Kin}_{k}[q^{-2} \hat{\bar{g}}].
\end{align}

However, usually we only consider the flow of somehow projected $n$-point functions $\hat{V}_{k, \lambda}^{(n)}$,
\begin{align}
 \hat{V}_{k, \lambda}^{(n)}[\hat{\bar{g}}] = \hat{\Gamma}_{k \, A_{1} \ldots A_{n}}^{(n)}[0;\hat{\bar{g}}]
 \mathbbm{P}_{\lambda}^{(n) \, A_{1} \ldots A_{n}}[\hat{\bar{g}}],
\end{align}
where $\mathbbm{P}_{\lambda}^{(n)}$ projects onto the physically most relevant or technically most feasible structures
of $\hat{\Gamma}_{k}^{(n)}$.
Here we use the index $\lambda$ to denote some tunable external parameter for the projection.
In practice this can, e.g., be an external momentum scale or numerating various tensor structures.
In order to derive the running of $\hat{V}_{k, \lambda}^{(n)}$,
\begin{align}
 \beta_{\hat{V}^{(n)}_{k, \lambda}}[\hat{\Gamma}_{k};\hat{R}, \hat{\bar{g}}]
 = \partial_{t} \hat{V}^{(n)}_{k, \lambda}[\hat{\bar{g}}],
\end{align}
we contract equation \eqref{eq:flow_Gamma_n} with $\mathbbm{P}_{\lambda}^{(n)}$,
\begin{widetext}
\begin{align} \label{eq:beta_V}
 \beta_{\hat{V}^{(n)}_{k, \lambda}}[\hat{\Gamma}_{k};\hat{R}, \hat{\bar{g}}]
 ={}& {\rm Flow}_{A_{1} \ldots A_{n}}^{(n)}[\hat{\bar{g}}] \mathbbm{P}^{(n) \, A_{1} \ldots A_{n}}_{\lambda}[\hat{\bar{g}}]
 + \partial_{q} \hat{V}^{(n)}_{k, \lambda}[q^{-2} \hat{\bar{g}}]
 \\ \notag
 {}&
 + \sum\limits_{l=1}^{n} \hat{\Gamma}_{k \, A_{1} \ldots A_{l-1} B A_{l+1} \ldots A_{n}}^{(n)}[0;\hat{\bar{g}}]
 \tensor{(\tfrac{1}{2} \eta_{k}[\hat{\bar{g}}] + \mathcal{N})}{^{B}_{A_{l}}}
 \mathbbm{P}_{\lambda}^{(n) \, A_{1} \ldots A_{n}}[\hat{\bar{g}}]
 - \hat{\Gamma}_{k \, A_{1} \ldots A_{n}}^{(n)}[0;\hat{\bar{g}}]
 \partial_{q} \mathbbm{P}^{(n) \, A_{1} \ldots A_{n}}_{\lambda}[q^{-2} \hat{\bar{g}}].
\end{align}
\end{widetext}
In some cases the terms in the second line of equation \eqref{eq:beta_V}
boil down to a simple factor multiplying $\hat{V}^{(n)}_{k, \lambda}$.
\subsection{Example: fermionic sector}
Let us be more explicit in the following.
As an example we consider again the fermionic sector from the main text, now in $d$-dimensional Euclidean space.
The fermionic terms we are following in our truncation can be written as
\begin{align} \label{eq:Gamma_Psi_Trunc_App}
 {}& \Gamma_{k}^{\psi} [h,\psi,\bar{\psi}; k^{-2} \delta]
 \\ \notag
 &{} = \! k^{\! -d} \scalebox{0.9}{$\sum\limits_{i=1}^{N_{\rm f}}$}
 \! \int \!\!\! \big[ k \bar{\psi}^{i} Z_{k}^{\psi}(\Box_{k}) \slashed{\partial} \psi^{i}
 \!\! - \! 2 \pi^{\! \frac{1}{2}} \! k^{3} \big( \! V_{k}(\Box_{k}) h^{{\scriptscriptstyle \rm TT}}_{\mu \nu} \big) \!
 \bar{\psi}^{i} \gamma^{\mu} \partial^{\nu} \! \psi^{i} \big] \! ,
\end{align}
where $\Box_{k} = k^{2} \Box$ is the dimensionful version of the dimensionless $\Box = - \partial^{2}$.
Here we already inserted a flat background, $k^{-2} \hat{\bar{g}}_{\mu \nu} = k^{-2} \delta_{\mu \nu}$.
The standard kinetic term for chiral fermions is
\begin{align}
 {\rm Kin}_{k}^{\psi}[k^{-2} \delta]_{i j}(x,y)
 = k^{1-d} \slashed{\partial} \delta(x-y) \delta_{i j},
\end{align}
where a factor $k^{-d}$ arises from the $\sqrt{\bar{g}}$ and one factor $k$ from the $\bar{\gamma}^{\mu}$
in the covariant formulation of the kinetic term,
$\int \sqrt{\bar{g}} \bar{\psi} \bar{\gamma}^{\mu} \bar{\nabla}_{\mu} \psi$.
Note that ${\rm Kin}_{k}^{\psi}$ corresponds to the $\psi$-$\bar{\psi}$ sector in field space,
\begin{align}
 {\rm Kin}_{k}[\bar{g}] \!
 = \!\! \begin{pmatrix}
 {\rm Kin}_{k}^{h}[\bar{g}] & 0 & 0 & 0 & 0 
 \\
 0 & 0 & \!\!\! {\rm Kin}_{k}^{\psi}[\bar{g}] & 0 & 0
 \\
 0 & \!\!\!\!\!\! - {\rm Kin}_{k}^{\psi \, \rm T}[\bar{g}] & 0 & 0 & 0
 \\
 0 & 0 & 0 & 0 & \!\!\! {\rm Kin}_{k}^{c}[\bar{g}]
 \\
 0 & 0 & 0 & \!\!\!\!\!\! - {\rm Kin}_{k}^{c \, \rm T}[\bar{g}] & 0
 \end{pmatrix} \! .
\end{align}
The generalized wave function renormalization $\mathcal{Z}_{k}^{\psi}$ then reads
\begin{align}
 \mathcal{Z}_{k}^{\psi}[k^{-2} \delta]_{i j}(x,y) = Z_{k}^{\psi \, \frac{1}{2}}(\Box_{k}) \delta(x-y) \delta_{i j}.
\end{align}
As we are dealing with chirally symmetric fermions there is
no explicit mass term, leading to an actually $k$-independent dimensionless kinetic operator $\widehat{\rm Kin}_{k}^{\psi}$,%
\footnote{%
If we dealt with nonchiral fermions we had
$\widehat{\rm Kin}^{\psi}_{k}[q^{-2} \delta]_{i j}(x,y)
= q^{-d} (q \slashed{\partial} + \hat{m}_{k}^{\psi}) \delta(x-y) \delta_{i j}$,
where $\hat{m}_{k}^{\psi}$ would be the dimensionless fermionic mass.
}
\begin{align}
 \widehat{\rm Kin}^{\psi}_{k}[q^{-2} \delta]_{i j}(x,y)
 = q^{1-d} \slashed{\partial} \delta(x-y) \delta_{i j}.
\end{align}
Here we can read off the canonical scaling $\mathcal{K}_{k}^{\psi}$
and the canonical mass dimension $\mathcal{N}_{k}^{\psi}$,
\begin{align}
 {}& \mathcal{K}_{k \, i j}^{\psi}(x,y) = k^{\frac{d-1}{2}} \delta(x-y) \delta_{i j},
 \\ \notag
 {}& \mathcal{N}_{k \, i j}^{\psi}(x,y) = \tfrac{d-1}{2} \delta(x-y) \delta_{i j}.
\end{align}
Thus the fermionic generalized anomalous dimension $\eta_{k}^{\psi}[\delta]$ is essentially
given by the standard fermionic anomalous dimension $\eta_{k}^{\psi}(\Box)$,
\begin{align}
 \eta^{\psi}_{k}[\delta]_{i j}(x,y)
 = - \frac{\dot{Z}^{\psi}_{k}(\Box_{k})}{Z^{\psi}_{k}(\Box_{k})} \delta(x \! - \! y) \delta_{i j}
 =  \eta^{\psi}_{k}(\Box) \delta(x \! - \! y) \delta_{i j}.
\end{align}
Putting everything together the dimensionless version of equation \eqref{eq:Gamma_Psi_Trunc_App} takes the form
\begin{align} \label{eq: gammakgen}
 {}& \hat{\Gamma}_{k}^{\psi} [\hat{h}, \hat{\psi}, \hat{\bar{\psi}}; q^{-2} \delta]
 \\ \notag
 {}& \hspace{0.3cm} = q^{-d} \sum\limits_{i = 1}^{N_{\rm f}}
 \! \int \! \big[ q \, \hat{\bar{\psi}}^{i} \slashed{\partial} \hat{\psi}^{i}
  - 2 \pi^{\frac{1}{2}} q^{3} \big( \hat{V}_{k}(\hat{\Box}_{q}) \hat{h}^{{\scriptscriptstyle \rm TT}}_{\mu \nu} \big)
 \hat{\bar{\psi}}^{i} \gamma^{\mu} \partial^{\nu} \psi^{i} \big] ,
\end{align}
where $\hat{\Box}_{q} = q^{2} \Box$ is the dimensionless metric scaled version of $\Box$.
To self-consistently determine the generalized anomalous dimension we plug the above
into equation \eqref{eq:flow_eta},
\begin{align}
 0 = {\rm Flow}_{\psi}^{(2)}[\delta]_{i j}(x,y)
 + \eta_{k}^{\psi}(\Box) \slashed{\partial} \delta(x \! - \! y) \delta_{i j}.
\end{align}
By first multiplying with $i \frac{\slashed{p}}{d_{\gamma} N_{\rm f} p^{2}} \delta(x) e^{i p \cdot y} \delta^{i j}$,
then taking the Dirac and flavor trace and finally integrating with respect to $x$ and $y$,
we arrive at
\begin{align}
 \eta_{k}^{\psi}(p^{2})
 = \int_{x, y} \!\! {\rm tr} \Big( i \frac{\slashed{p}}{d_{\gamma} N_{\rm f} p^{2}} e^{i p \cdot y} \delta(x) \delta^{i j}
 {\rm Flow}_{\psi}^{(2)}[\delta]_{i j}(x,y) \Big).
\end{align}
This projection can be summarized as
\begin{align}
 \mathbbm{P}^{(2)}_{p}[q^{-2} \delta]^{i j}(x,y)
 = i \, q^{d-1} \, \frac{\slashed{p}}{d_{\gamma} N_{\rm f} \, p^{2}} \delta(x) e^{i p \cdot y} \delta^{i j}.
 \label{eq: Proj2}
\end{align}
Finally let us move on to the running of $\hat{V}_{k}$.
The full graviton-fermion three-point function depends on three coordinates $x$, $y$ and $z$,
and therefore on three external momenta in Fourier space, $p_1$, $p_2$ and $p_3$.
One of them can be eliminated due to momentum conservation.
As projector $\mathbbm{P}_{p_{2},p_{3}}^{(3)}$ on the desired quantity $\hat{V}_{k}$, we use
\begin{align} \notag
 \mathbbm{P}_{p_{2},p_{3}}^{(3)} {}& [q^{-2} \delta]_{\mu \nu}^{i j}(x,y,z)
 \\
 {}& \hspace{-0.9cm}
 = \! \frac{i q^{d-3} 8 (d \! - \! 1) \gamma_{\rho} \, p_{2\sigma}}{
 6 \pi^{\frac{1}{2}} d_{\gamma} N_{\rm f} (d \! + \! 1)(d \! - \! 2) p^{2}}
 \Pi^{{\scriptscriptstyle \rm TT} \,\, \rho \sigma}_{\, \mu\nu} \! (p_{1})
 e^{i ( p_{2} \cdot y + p_{3} \cdot z )}
 \delta(x) \delta^{i j},
 \label{eq: Proj3_App}
\end{align}
which we evaluate at the momentum symmetric point,
where $p_{2}^{2} = p_{3}^{2} = - 2 (p_{2} \cdot p_{3}) = p^{2}$ and $p_{1} = -p_{2} - p_{3}$.
This is the $d$-dimensional version of Eq.~\eqref{eq: Proj3}.
The normalization of $\mathbbm{P}^{(3)}_{p_{2},p_{3}}$ again follows from
\begin{align}
 \Pi^{{\scriptscriptstyle \rm TT} \, \, \mu \nu}_{\, \mu \sigma}(p_{1})
 p_{2}^{\sigma} p_{2 \, \nu}
 = \frac{(d+1)(d-2)}{2(d-1)} \Big( p_{2}^{2} - \tfrac{(p_{1} \cdot p_{2})^{2}}{p_{1}^{2}} \Big),
\end{align}
when evaluated at the momentum symmetric point, $p_{1}^{2} = p_{2}^{2} = -2 (p_{1} \cdot p_{2}) = p^{2}$.
Using the general expression \eqref{eq:beta_V} for $\beta_{\hat{V}^{(n)}_{k,\lambda}}$
and inserting the projector \eqref{eq: Proj3_App} we finally arrive at the $d$-dimensional version for $\beta_{\hat{V}_{k}}$,
cf.~\eqref{eq: vertflow},
\begin{align}
 \beta_{\hat{V}_{k}}(p^{2})
 ={}& \mathrm{Flow}^{(3)}_{\psi}(p^2)\notag
 \\
 {}& \!\! + \hat{V}_{k}(p^{2})
 \big( \tfrac{d-2}{2} \! + \! \eta_{k}^{\psi}(p^{2}) \! + \! \tfrac{1}{2} \eta_{k}^{h}(p^{2}) \big)
 \! + \! 2 p^{2} \hat{V}_{k}'(p^{2})\,.
\end{align}
\section{Beta functions for avatars of the Newton coupling from the derivative expansion}\label{sec:appderbeta}
Extracting all beta functions and anomalous dimensions from the derivative expansion and then equating $G_{\psi}=G=G_h$
results in the following two beta functions for the avatars of the Newton coupling.
We highlight the last term in each expression, which encodes the effect of fermionic fluctuations.
These screen the gravitational couplings for $\sigma>-0.07$.
\pagebreak
\begin{widetext}
\noindent
Here and in the following $\PTT=\frac{1}{1-2\lambda_2}$ and $\PTr=\frac{1}{3-4\lambda_2}$ denote the pole structure of
the TT-and trace mode, respectively.
\begin{align}
 \beta_{G_h}
 =&2G+\frac{G^2}{\pi}\bigg(-\frac{27}{95}\PTr^5+\frac{175}{152}\PTr^4+9\PTr^2-\frac{85}{342}\PTT^2
 -\frac{1}{5}\PTr^3\PTT^2+\frac{54}{95}\PTr^4\PTT\notag
 \\
 &+\frac{277}{2280}\PTr^3\PTT+\frac{2(-439+932\lambda_3)}{855}\PTr\PTT^4-\frac{5(-3265+1152\lambda_3)}{4104}\PTr\PTT^2
 \notag
 \\
 &+\frac{5(1069+1728\lambda_3)}{1368}\PTr^2\PTT+\frac{(-1811+3508\lambda_3)}{285}\PTr^2\PTT^3
 \notag
 \\
 &-\frac{23(-13263+39472\lambda_3)}{61560}\PTr\PTT^3+\frac{(-136953+231368\lambda_3)}{20520}\PTr^2\PTT^2
 \notag
 \\
 &+\frac{(-21565+47432\lambda_3+30664\lambda_3^2)}{5130}\PTT^3
 +\frac{(-2689+15948\lambda_3-33144\lambda_3^2+20768\lambda_3^3)}{2565}\PTT^5
 \notag
 \\
 &+\frac{(251791-1241168\lambda_3+1676088\lambda_3^2+2606848\lambda_3^3)}{61560}\PTT^4-\frac{287}{4104}
 +N_{\mathrm{f}}\frac{19439+165440\sigma}{159600}\bigg)+\mathcal{O}(G^3)\,,
 \label{eq: GhDeriv}
\end{align}
\begin{align}
 \beta_{G_{\psi}}
 =&2G+\frac{G^2}{\pi}\bigg(\frac{3}{8}\PTr^4-\frac{25}{12}\PTr^2\PTT^2+\frac{5(71-272\lambda_3
 +296\lambda_3^2)}{216}\PTT^4+\frac{4(-9+20\sigma^2)}{81}\PTT
 \notag
 \\
 &-\frac{10(-11+28\sigma^2)}{231}\PTr\PTT+\frac{2(-81+180\sigma+80\sigma^2)}{81}\PTT^2
 +\frac{27+112\sigma+252\sigma^2}{42}\PTr^3
 \notag
 \\
 &-\frac{5(-3135+1232\sigma+2688\sigma^2)}{5544}\PTr\PTT^2-\frac{5(253+1232\sigma+2688\sigma^2)}{1848}\PTr^2\PTT
 \notag
 \\
 &+\frac{4536+9045\sigma+7168\sigma^2}{7560}\PTr-\frac{12528+73855\sigma+100072\sigma^2}{2520}\PTr^2
 \notag
 \\
 &+\frac{5(-17+24\lambda_3^2-96\sigma+80\lambda_3(-3+2\sigma))}{162}\PTT^3-\frac{1}{324}+N_{\mathrm{f}}
 \frac{105+1472\sigma}{720}\bigg)+\mathcal{O}(G^3)\,,
 \label{eq: GpsiDeriv}
\end{align}
\begin{align}
 \beta_{\sigma}=&2\sigma+\frac{G}{\pi}\bigg(-\frac{25(-1+2\sigma+6\sigma^2)}{24}\PTT^2\PTr^2
 +\frac{3(27+140\sigma+300\sigma^2)}{256}\PTr^4-\frac{\sigma(2268-9999\sigma+896\sigma^2)}{12096}\PTr
 \notag
 \\
 &-\frac{1155-12012\sigma+17370\sigma^2+8960\sigma^3}{18144}\PTT
 +\frac{1386-12375\sigma+10010\sigma^2+10080\sigma^3}{16632}\PTT\PTr
 \notag
 \\
 &+\frac{5082+4125\sigma+29260\sigma^2+20160\sigma^3}{16632}\PTr\PTT^2-\frac{13041+85104\sigma+134428\sigma^2
 +48384\sigma^3}{16128}\PTr^3
 \notag
 \\
 &+\frac{20979+128628\sigma+243065\sigma^2+236544\sigma^3}{12096}\PTr^2+\frac{5(51-74\sigma
 +\lambda_3(-111+124\sigma))}{432}\PTT^4
 \notag
 \\
 &-\frac{6741+6858\sigma+306590\sigma^2+53760\sigma^3-300\lambda_3(49-126\sigma+96\sigma^2)}{54432}\PTT^2
 \notag
 \\
 &+\frac{6237+11055\sigma-7700\sigma^2+20160\sigma^3}{5544}\PTr^2\PTT
 \notag
 \\
 &-\frac{1386-9690\sigma-21280\sigma^2+15\lambda_3(-749-2730\sigma+740\sigma^2)}{13608}\PTT^3\bigg)+\mathcal{O}(G^2).
 \label{eq:appbetasigexp}
\end{align}
In the following we list the general beta function for the fermion-gravity avatar of the Newton coupling $G_{\psi}$
and the nonminimal coupling $\sigma$, obtained via a derivative expansion:
\begin{align}
 \beta_{G_{\psi}}=&(2+\eta_{h_{\rm TT}}+2\eta_{\psi})G_{\psi}+\frac{G_{\psi}^2}{\pi}\bigg(-\frac{2(18-40\sigma^2
 +\eta_{\psi}(-3+4\sigma^2))}{81}\PTT
 \notag
 \\
 &+\frac{154(-891+5040\sigma+1280\sigma^2)+\eta_{h_{\rm TT}}(-207575+97020\sigma+17920\sigma^2)}{99792}\PTT^2
 \notag
 \\
 &+\frac{-88(3591+58095\sigma+66556\sigma^2)\eta_{h_{\rm Tr}}(62469+623172\sigma+552104\sigma^2)}{221760}
 \notag
 \\
 &+\frac{\eta_{\psi}(4158+9405\sigma+3304\sigma^2)-11(2268+7155\sigma+3416\sigma^2)}{83160}\PTr\bigg)
 \notag
 \\
 &+\frac{\sqrt{G_{\psi}}G_h^{\tfrac{3}{2}}}{\pi}\bigg(\frac{5(264-672\sigma^2
 +\eta_{\psi}(-33+56\sigma^2))}{2772}\PTr\PTT+
 \frac{\eta_{\psi}(81+224\sigma+168\sigma^2)-8(81+280\sigma+252\sigma^2)}{2016}\PTr^2
 \notag
 \\
 &+\frac{-520(-121+231\sigma+504\sigma^2)+3\eta_{h_{\rm TT}}(-2145+4004\sigma+6720\sigma^2)}{108108}
 \big(\PTr\PTT^2+3\PTr^2\PTT\big)
 \notag
 \\
 &-\frac{-715(27+112\sigma+252\sigma^2)+\eta_{h_{\rm Tr}}(2145+6734\sigma+13680\sigma^2)}{30030}\PTr^3\bigg),
\end{align}
\begin{align}
 \beta_{\sigma}
 =&2\sigma+\frac{\sqrt{G_hG_{\psi}}}{\pi}
 \bigg(-\frac{25(-1+2\sigma+6\sigma^2)}{24}\PTr^2\PTT^2+\frac{3(27+140\sigma+300\sigma^2)}{256}\PTr^4
 \notag
 \\
 &+\frac{5(51-74\sigma+\lambda_3(-111+124\sigma))}{432}\PTT^4
 +\frac{5\lambda_3\eta_{h_{\mathrm{TT}}}(-861-3030\sigma+68\sigma^2)}{27216}\PTT^3
 \notag
 \\
 &-\frac{\eta_{\psi}(1848-12375\sigma+8008\sigma^2+6720\sigma^3)-8(1386-12375\sigma+10010\sigma^2+10080\sigma^3)}{
 133056}\PTr\PTT
 \notag
 \\
 &-\frac{-312(6237+11055\sigma-7700\sigma^2+20160\sigma^3)+\eta_{h_{\mathrm{Tr}}}(59202+368225\sigma+299208\sigma^2
 +482840\sigma^3)}{1729728}\PTr^2\PTT
 \notag
 \\
 &+\frac{\eta_{h_{\mathrm{TT}}}(10692-54175\sigma-93184\sigma^2)+88(-693+4845\sigma+10640\sigma^2)
 +660\lambda_3(749+2730\sigma-740\sigma^2)}{598752}\PTT^3
 \notag
 \\
 &-\frac{-2\eta_{\psi}(63-225\sigma+196\sigma^2)+4(189-900\sigma+980\sigma^2)+75\lambda_3
 (\eta_{\psi}(49-84\sigma+48\sigma^2)-4(49-126\sigma+96\sigma^2))}{54432}\PTT^2
 \notag
 \\
 &-\frac{\eta_{h_{\mathrm{Tr}}}(521235+471900\sigma-2475836\sigma^2-5322240\sigma^3)}{23063040}\PTr^3
 \notag
 \\
 &-\frac{\sigma(\eta_{\psi}(81+280\sigma+252\sigma^2)-8(81+350\sigma+378\sigma^2))}{6048}\PTr^2
 \notag
 \\
 &+\frac{567\eta_{\psi}(1+2\sigma)^2-2(1701+12528\sigma+34636\sigma^2+48384\sigma^3)}{32256}\PTr^3\bigg)
 \notag
 \\
 &+\frac{G_{\psi}}{\pi}\bigg(\frac{\eta_{\psi}(609-4044\sigma+4360\sigma^2+1792\sigma^3)-2(1155-12012\sigma
 +17370\sigma^2+8960\sigma^3)}{36288}\PTT
 \notag
 \\
 &-\frac{33(1995+3486\sigma+100890\sigma^2+17920\sigma^3)-3\eta_{h_{\mathrm{TT}}}(1386+13728\sigma+143407\sigma^2
 +17920\sigma^3)}{598752}\PTT^2
 \notag
 \\
 &-\frac{\eta_{h_{\mathrm{Tr}}}(189189+766755\sigma+971685\sigma^2+768320\sigma^3)}{443520}\PTr^2
 \notag
 \\
 &-\frac{189\eta_{\psi}(5+32\sigma+44\sigma^2)-4(6993+42462\sigma+79155\sigma^2+76832\sigma^3)}{16128}\PTr^2
 -\frac{9(5+32\sigma+44\sigma^2)}{64}\PTr^3\bigg).
 \label{eq:appbetasig}
\end{align}
\end{widetext}

\bibliography{refs_v2}

\end{document}